\documentclass[aps,preprint,prd,nofootinbib,superscriptaddress,
]{revtex4-2}	
\pdfoutput=1	
\usepackage[utf8]{inputenc}
\usepackage{booktabs}
\usepackage{amssymb,amsmath,amsfonts}
\usepackage{slashed}
\usepackage{indentfirst}
\usepackage[pdftex]{graphicx}
\usepackage{epstopdf}
\usepackage{psfrag}

\usepackage{float,multirow,rotfloat,supertabular} 
\usepackage{hyperref}
\usepackage{mathrsfs}
\usepackage{url}
\usepackage{appendix}
\usepackage{subfigure}
\usepackage{color}
\usepackage[dvipsnames]{xcolor}
\usepackage{soul}
\usepackage{bm}
\usepackage{siunitx}
\usepackage{gensymb} 

\usepackage{array}
\usepackage{cancel}
\usepackage[compat=1.1.0]{tikz-feynman}

\usepackage{multirow}
\usepackage{url}
\usepackage{adjustbox}

\DeclareUnicodeCharacter{2212}{-} 

\allowdisplaybreaks	

\newcommand{\TeV}{\text{ TeV}}

\newcommand{\ab}{\text{ ab}}

\graphicspath{{./figures/}}

\def\figureautorefname~#1\null{Fig.\,#1\null}

\def\equationautorefname~#1\null{Eq.\,(#1)\null}

\newcommand{\bpm}{\begin{pmatrix}}
\newcommand{\epm}{\end{pmatrix}}

\newcommand{\beq}{\begin{equation} }
\newcommand{\eeq}{\end{equation} }

\newcommand{\inab}{\,{\rm ab}^{-1}}

\newcommand{\cosbt}{|\cos{\bar{\theta}}|}

\begin{document}
\count\footins = 750    

\title{Probing Z/W Pole Physics at High-energy Muon Colliders via Vector-boson-fusion Processes}

\author{Hao-Qiao Li}
\email{hqli22@m.fudan.edu.cn}
\affiliation{Department of Physics and Center for Field Theory and Particle Physics, Fudan University, Shanghai 200438, China}

\author{Hai-Ning Yan}
\email{hnyan22@m.fudan.edu.cn}
\affiliation{Department of Physics and Center for Field Theory and Particle Physics, Fudan University, Shanghai 200438, China}

\author{Jiayin Gu}
\email{jiayin\_gu@fudan.edu.cn}
\affiliation{Department of Physics and Center for Field Theory and Particle Physics, Fudan University, Shanghai 200438, China}
\affiliation{Key Laboratory of Nuclear Physics and Ion-beam Application (MOE), Fudan University, 
Shanghai 200433, China}

\author{Xiao-Ze Tan}
\email{xz\_tan@fudan.edu.cn}
\affiliation{Department of Physics and Center for Field Theory and Particle Physics, Fudan University, Shanghai 200438, China}
\affiliation{Deutsches Elektronen-Synchrotron DESY, Notkestr.\ 85, 22607 Hamburg, Germany}

\begin{abstract}

A future $e^+e^-$ collider could run at the Z-pole to perform important electroweak (EW) precision measurements, while such a run may not be viable for a future muon collider.  This however can be compensated by the measurements of other EW processes, taking advantage of the high energy and large luminosity of the muon collider.  In this paper, we consider the measurements of the vector boson fusion processes of $WW/WZ/W\gamma$ to a pair of fermions (along with a $\nu_{\mu}\bar{\nu}_{\mu}$ or $\nu_{\mu}\mu^+/\bar{\nu}_{\mu}\mu^-$ pair) at a high-energy muon collider and study their potential in probing the EW observables.  We consider two run scenarios for the muon collider with center-of-mass energy of 10\,TeV and 30\,TeV, respectively, and focus on the processes involving $f=b,c,\tau$ and 
the dimension-6 operators that directly modify the corresponding fermions coupling to the $Z/W$ bosons. 
The invariant mass distribution of the $f\bar{f}$ pair helps to separate the events from the $Z/W$ resonance and the high-energy ones, while the polar angle of the outing fermion also provides additional information.     
By performing a chi-squared analysis on the binned distributions and combining the information from the $WW$ and $WZ/W\gamma$ fusion processes, all relevant Wilson coefficients can be constrained simultaneously. The precision surpasses the current EW measurement constraints and is even competitive with future $e^+e^-$ colliders. 
Our analysis can be included in a more complete framework which is needed to fully determine the potential of muon colliders in EW precision measurements.

\end{abstract}

\preprint{DESY-25-051}

\maketitle

\newpage
\tableofcontents

\newpage
\section{Introduction}

A future high-energy muon collider, if constructed, will have enormous physics potential and may unveil some of the deepest mysteries in particle physics. Despite significant technical challenges, the concept of a muon collider was proposed in the early days of particle physics and has recently gained strong interest within the global scientific community~\cite{AlAli:2021let,Aime:2022flm, Accettura:2023ked}.  
The Muon Accelerator Program (MAP), initiated in the United States, put forward a concrete framework for developing a muon collider, with an envisioned startup timeline around 2045~\cite{Narain:2022qud}. Meanwhile, the European Strategy for Particle Physics (ESPPU), updated in 2020, emphasized the importance of a robust R\&D program for muon colliders~\cite{EuropeanStrategyGroup:2020pow}. This initiative led to the establishment of the International Muon Collider Collaboration (IMCC)~\cite{IMCC}, tasked with delivering a promising Conceptual Design Report (CDR) in time for the next ESPPU update~\cite{InternationalMuonCollider:2024jyv}. Additionally, the European Roadmap for Accelerator R\&D~\cite{Adolphsen:2022ibf}, published in 2021, included the muon collider as a priority, based on recommendations from the particle physics community. 
This roadmap outlined critical requirements for key technologies and conceptual advances, setting benchmarks for technical assessments such as luminosity goals and the management of detector backgrounds~\cite{Accettura:2023ked}. 
Over the coming decades, studies on muon colliders are expected to play an increasingly pivotal role in particle physics, offering unique opportunities to explore physics at unprecedented energy scales. 

The main advantage of a muon collider, compared with an $e^+e^-$ collider (such as the CEPC~\cite{CEPCStudyGroup:2018ghi, CEPCPhysicsStudyGroup:2022uwl}, FCC-ee~\cite{FCC:2018evy, Bernardi:2022hny}, ILC~\cite{ILCInternationalDevelopmentTeam:2022izu}, C$^3$~\cite{Bai:2021rdg} and CLIC~\cite{CLICdp:2018cto, CLIC:2018fvx}), is that it can reach a higher center-of-mass energy, possibly up to 10\,TeV or even 30\,TeV~\cite{Aime:2022flm, Accettura:2023ked}, thanks to the larger muon mass which greatly reduces the synchrotron radiation effects.  
A possible 3 TeV run is also under consideration~\cite{MuonCollider:2022xlm}, while a 125\,GeV run as a first stage to produce Higgs boson on $s$-channel resonance could still be strategically advantageous from both physics and accelerator perspectives~\cite{Barger:1995hr,Barger:1996jm,deBlas:2022aow}. 
With the advantage of reaching the highest possible energy and directly searching for new physics beyond the Standard Model (BSM), a muon collider does fall short on the electroweak (EW) precision program due to the lack of a Z-pole run, as opposed to an $e^+e^-$ collider.  

However, all is not lost.  The lack of a Z-pole program can be made up by exploiting other EW measurements.  This is manifest in the framework of the Standard Model Effective Field Theory (SMEFT), where the new physics effects are systematically parametrized in terms of a series of higher dimensional operators, 
and the connections among different couplings imposed by the SM gauge groups are automatically imposed.
Two important aspects should be considered for the EW analysis under the SMEFT framework.  The first is that the contributions of a large set of higher dimensional operators have energy enhancements, and it is advantageous to measure the corresponding processes at the highest possible energy, despite a possible smaller SM cross section.  A typical example is the diboson ($WW/WZ/Vh$) processes, as their measurements at a high energy muon collider can have competitive or even better reaches on certain EW operators compared with the future $Z$-pole program~\cite{Franceschini:2017xkh, Buttazzo:2020uzc}.  However, the diboson production is not directly sensitive to the fermion couplings other than the ones of muon, while the subsequent decays of $W$ or $Z$ (which do depend on fermion couplings) do not have energy enhancements.  The second aspect is that the cross sections of vector boson fusion (VBF) processes increase with the collider energy, and a large number of events can be collected at the high energy runs, effectively turning the high energy into high precision~\cite{Buttazzo:2020uzc, Han:2020uid, Han:2020pif,Costantini:2020stv,Forslund:2022xjq,Han:2023njx}.  

In this paper, we further investigate the potential of EW measurements at a high energy muon collider by performing phenomenological analyses to the fusion processes of two vector bosons (with at least one $W$ boson, {\it i.e.} $WW/WZ/W\gamma$) into a pair of fermions (along with a $\nu_{\mu}\bar{\nu}_{\mu}$ or $\nu_{\mu}\mu^+/\bar{\nu}_{\mu}\mu^-$ pair coming from the incoming $\mu^+\mu^-$), which we collectively denote as VBF$\to 2f$.\footnote{
The VBF$\to 2f$ processes considered here do not necessarily involve an $s$-channel W/Z boson, in which case the use of the term ``fusion'' is somewhat inaccurate.  For convenience, we simply use the term ``VBF$\to 2f$'' throughout this paper to denote the two-vector-boson-to-two-fermion processes, with at least one of the two vector bosons being $W$.  We also decided to avoid using the term ``vector boson scattering'' (VBS) which is often reserved for $VV\to VV$ processes.    
} 
We focus on the cases with either bottom, charm or tau in the final states, 
and consider the operators that directly modify their couplings to the $W/Z$ bosons.  As mentioned above, these operators cannot be directly probed by the diboson production processes without additional flavor assumptions, and the measurements of the VBF processes thus provide valuable complementary information.  In addition, the invariant mass distribution of the fermion pair contains useful information, which can be extracted with a binned analysis.\footnote{For the $WZ/W\gamma$ fusion processes with lepton final states, the invariant mass can not be reconstructed due to the missing neutrino.}  
By combining the measurements of different fusion processes and exploiting the differential information, all relevant operator coefficients can be constrained simultaneously.  In some cases, the reaches are even competitive with those of future $e^+e^-$ colliders.    

The rest of this paper is organized as follows: In \autoref{sec:theory}, we provide an overview of the relevant dimension-6 SMEFT operators and the details of the VBF$\to 2f$ processes with a discussion on their characteristic features.  
Our results are provided in \autoref{sec:results}, which include detailed comparisons that illustrate the importance of the differential information and the complementarity of the $WW$ and $WZ/W\gamma$ fusion processes.  The overall reaches of the Wilson coefficients from our analysis are summarized in \autoref{fig:delta}. 
Finally, the conclusion is drawn in \autoref{sec:con}.
The numerical expressions for the binned cross sections are provided in \ref{app:xs} for all the processes included in our analysis.  Additional results are provided in \ref{app:delta_rho}.

\section{The theoretical framework}
\label{sec:theory}

\subsection{Dimension-6 operators in SMEFT}
\label{subsec:d6}

The SMEFT Lagrangian can be obtained by adding to the SM Lagrangian a series of higher dimensional operators, characterized by the energy scale $\Lambda$~\cite{Buchmuller:1985jz, Grzadkowski:2010es}.  Assuming baryon and lepton numbers are conserved around the TeV scale, all higher dimensional operators are of even dimensions, 
\begin{equation}
\mathcal{L}_{\rm SMEFT} =
    \mathcal{L}_{\rm SM} + 
    \sum_i \frac{c^{(6)}_i}{\Lambda^2} \mathcal{O}^{(6)}_i + 
    \sum_j \frac{c^{(8)}_j}{\Lambda^4} \mathcal{O}^{(8)}_j +         \cdots   \label{eq:smeft}
\end{equation}
where the leading new physics contributions are given by the dimension-6 operators. 
The phenomenology of the SMEFT dimension-6 operators is an active field, with many global analyses already performed for the EW, Higgs and top measurements at both current and future colliders~\cite{Falkowski:2015jaa, Durieux:2017rsg, Barklow:2017suo, Durieux:2018tev, Durieux:2018ggn, Ellis:2018gqa, DeBlas:2019qco, Durieux:2019rbz, Ellis:2020unq, Miralles:2021dyw,Ethier:2021bye,Bruggisser:2022rhb,Liu:2022vgo, deBlas:2022ofj,  Brivio:2022hrb, Aoude:2022aro,Bartocci:2023nvp, Allwicher:2023shc, Grunwald:2023nli, Garosi:2023yxg,Elmer:2023wtr,Kassabov:2023hbm,Bellafronte:2023amz,Celada:2024mcf,Asteriadis:2024xuk,Bartocci:2024fmm,ElFaham:2024egs,ElFaham:2024uop,Celada:2024mcf,
Biekotter:2025nln,Cornet-Gomez:2025jot,terHoeve:2025gey,Maura:2025rcv}. 
A similar global analysis that includes all the EW measurements at a muon collider should also be performed to fully understand the physics potential.  However, such an analysis is beyond the scope of our current study.  Instead, we focus on the set of operators that directly modifies the couplings of fermions to the gauge bosons ($Vff$-type couplings), which are   
\begin{align} 
     \mathcal{O}^{(1)}_{Hl}  = {}& (H^{\dagger} i \overleftrightarrow{D}_\mu H)(\bar{\ell} \gamma^\mu \ell) \,, \nonumber\\
    \mathcal{O}^{(3)}_{Hl}  = {}& (H^{\dagger} i \overleftrightarrow{D}_\mu^i H)(\bar{\ell} \sigma^i \gamma^\mu \ell) \,, \nonumber\\
    \mathcal{O}_{He} = {}& (H^{\dagger} i \overleftrightarrow{D}_\mu  H)(\bar{e} \gamma^\mu e) \,,  \label{eq:Op1}
\end{align}    
and similar ones for the quarks, 
\begin{align} 
     \nonumber\\
    \mathcal{O}^{(3)}_{Hq} = {}& (H^{\dagger} i \overleftrightarrow{D}_\mu^i  H)(\bar{q} \sigma^i \gamma^\mu q) \,, \nonumber\\
    \mathcal{O}_{Hu} = {}& (H^{\dagger} i \overleftrightarrow{D}_\mu  H)(\bar{u} \gamma^\mu u) \,, \nonumber\\
    \mathcal{O}_{Hd} = {}& (H^{\dagger} i \overleftrightarrow{D}_\mu  H)(\bar{d} \gamma^\mu d) \,, \label{eq:Op2}
\end{align}
where
\begin{align}
    H^{\dagger} i \overleftrightarrow{D}_\mu H = {}& H^{\dagger}(i D_\mu H) - (i D_\mu H^\dagger) H, \\
    H^{\dagger} i \overleftrightarrow{D}_\mu^i H = {}& H^{\dagger} \sigma^i (i D_\mu H) - (i D_\mu H^\dagger) \sigma^i H \,,
\end{align}
and the flavor/generation indices are not explicitly written out.  
After EWSB, they generate modifications to the SM $Vff$-type couplings (as well as $hVff$ contact interactions which we do not consider in this study) which can be parameterized as 
\begin{align} 
\mathcal{L} \supset &~~ - \frac{g}{c_W} Z_\mu \left[ \sum_{f=u,d,e,v}\!\!\!  \bar{f}_L \gamma^\mu (T^3_f - s^2_W Q_f +  \delta g^{Zf}_L ) f_L + \!\!\!  \sum_{f=u,d,e} \!\!\!\bar{f}_R \gamma^\mu (- s^2_W Q_f + \delta g^{Zf}_R) f_R    \right]  \nonumber\\
 & ~ -  \frac{g}{\sqrt{2}} \left[  W^+_\mu \bar{u}_L \gamma^\mu ( V_{\rm CKM} + \delta g^{Wq}_L ) d_L + W^+_\mu \bar{\nu}_L \gamma^\mu (1+ \delta g^{W\ell}_L ) e_L + {\rm h.c.}   \right]  \,,  \label{eq:Lew}
\end{align}
where $u,d,e$ collectively denote the corresponding fermions of all 3 generations.  Assuming $V_{\rm CKM} =1$ for simplicity, the modifications of the $Vff$-type couplings are related to the Wilson coefficients by 
\begin{align}
    \delta g_L^{Ze}=&\, -(c_{H \ell}^{(1)}+ c_{H \ell}^{(3)})  \frac{v^{2} }{2 \Lambda^{2} } \,, 
    \hspace{1cm} \delta g_R^{Ze}=-c_{H e}	\frac{v^{2} }{ 2\Lambda^{2} } \,,  
    \hspace{1cm} \delta g_L^{Wl} = c_{H l}^{(3)} \frac{v^{2} }{ \Lambda^{2} } \,, 
    \nonumber \\
    \delta g_L^{Zu}=&\, -(c_{H q}^{(1)}- c_{H q}^{(3)})  \frac{v^{2} }{2 \Lambda^{2} } \,, 
    \hspace{1cm} \delta g_R^{Zu}= -c_{H u}	\frac{v^{2} }{ 2\Lambda^{2} } \,, 
    \nonumber \\
	\delta g_L^{Zd}=&\, -(c_{H q}^{(1)}+ c_{H q}^{(3)})  \frac{v^{2} }{2 \Lambda^{2} } \,, 
    \hspace{1cm} \delta g_R^{Zd}= -c_{H d}	\frac{v^{2} }{ 2\Lambda^{2} } \,, 
    \hspace{1cm} \delta g_L^{Wq} = c_{H q}^{(3)} \frac{v^{2} }{ \Lambda^{2} } \,,
\label{eq:dg2c}
\end{align}
where again the generation indices are not explicitly written.

We assume that all operators/couplings are flavor diagonal and focus on those that contribute to the processes involving $b,\,c$ or $\tau$ (as listed later in \autoref{subsec:proc}), while the couplings of different generations are allowed to be different ({\it i.e.,} flavor universality is not imposed).  The reasons for our choices are as follows. 
First, operators that generate universal corrections tend to be better constrained by other measurements, such as the $W$ mass measurement, or the measurement of the diboson or $Vh$ processes which have a much larger (effective) c.o.m. energy~\cite{Buttazzo:2020uzc}.  Assuming this is the case, it is then a reasonable approximation to set their effects to zero for the VBF$\to 2f$ processes.  The same applies for the EW operators that involve the electron or the muon.  For instance, there is no need to consider the process $ W^+W^- \to \mu^+\mu^-$ when the reverse one, $\mu^+\mu^-\to W^+W^-$, is available with a much larger c.o.m. energy.  Processes involving only the light quarks ($u,d,s$) are difficult to tag, and the corresponding operators cannot be well constrained without additional flavor assumptions.  Finally, the processes involving the top quark are of crucial importance but are left for future studies since the kinematics are much more complicated due to the top decay.  

Throughout most of this paper, we will omit the generation indices of the Wilson coefficients for simplicity, which is hopefully clear from the content.  (Note again that we do not assume flavor universality in our study.)
For the bounds summarized in \autoref{fig:delta}, the generations are denoted by subscripts $1,2,3$.

\subsection{ Processes}
\label{subsec:proc}

\begin{figure}[htb]
    \centering
	\includegraphics[width=0.70\textwidth]{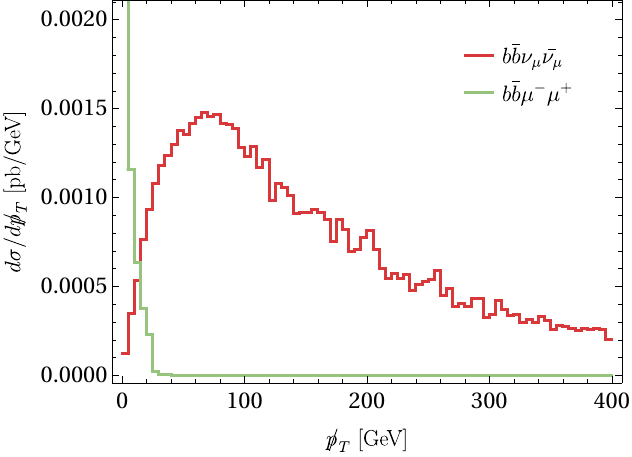}
\caption{
The missing transverse momentum ($\slashed{p}_T$) distribution for $b\bar{b}\nu_\mu\bar{\nu}_\mu$ (red) and $b\bar{b}\mu^-\mu^+$ (green) at $\sqrt{s}=10\,$TeV. For $b\bar{b}\mu^-\mu^+$, a cut on the muon rapidity of $|\eta_\mu|>6$ is imposed, corresponding to both muons being untagged.   
}
\label{fig:pt_dis_bb_10}
\end{figure}

We consider the vector boson fusion processes that produce a pair of fermions (VBF$\to 2f$) where at least one of the fermions is either $b$, $c$ or $\tau$. More explicitly, including all final state particles, we consider the following five $\mu^-\mu^+ \to 4f$ processes: 
\begin{align}
\mu^-\mu^+ \to b \bar{b}\nu_{\mu}\bar{\nu}_{\mu}\,,\hspace{1cm} &\mu^-\mu^+ \to c \bar{c}\nu_{\mu}\bar{\nu}_{\mu}\,,\hspace{1cm} \mu^-\mu^+ \to \tau^- \tau^+\nu_{\mu}\bar{\nu}_{\mu}\,, \nonumber\\
& \hspace{-1cm} \mu^-\mu^+ \to c s \nu_{\mu} \mu \,,\hspace{1cm} \mu^-\mu^+ \to\tau \nu_\tau \nu_{\mu} \mu \,,
\end{align}
where by $c s \nu_{\mu} \mu$ we denote the combination of $c\bar{s} \bar{\nu}_\mu\mu^-$ and $\bar{c}s \nu_\mu\mu^+$, and the same applies to $\tau \nu_\tau \nu_{\mu} \mu$.  
Note that the $\mu^-\mu^+ \to t b \nu_{\mu} \mu$ process is not considered in our analysis since the top decay produces more complicated kinematic features, which requires separate studies. 
Here we shall take the $\mu^-\mu^+ \to b \bar{b}\nu_{\mu}\bar{\nu}_{\mu}$ process as an example, noting that the following discussion also applies to the other processes to a large extent.  
As shown in \autoref{fig:FDbb}, many diagrams contribute to the $4f$ final states.  However, the cross section is dominated by the $WW$ fusion process, especially when the invariant mass of the $b\bar{b}$ pair is around the $Z$-pole.
For instance, at 10\,TeV the total tree-level cross section of this process is 0.30\,pb, while the one of $\mu^-\mu^+ \to Z\bar{\nu}_{\mu} \nu_{\mu} \,,~Z\to  b \bar{b}$ 
is around 0.29\,pb %
according to MadGraph5~\cite{Alwall:2011uj}.  
Furthermore, the diboson process $\mu^-\mu^+ \to ZZ \to \bar{\nu}_{\mu} \nu_\mu  b \bar{b}$ has a much smaller cross section, which is around $65\,$ab.   Similarly, the irreducible backgrounds $\mu^-\mu^+ \to ZZ \to \bar{\nu}_{e} \nu_e  b \bar{b}/\bar{\nu}_{\tau} \nu_\tau  b \bar{b}$ also have negligible rates compared with the signal.  
Note that in our analysis, we always simulate the full $\mu^-\mu^+ \to 4f$ process which includes all diagrams except those that involves the Higgs boson, which can be easily removed with an invariant mass window cut around the Higgs mass.  
The signal events can be selected in experiments by requiring a pair of fermions along with missing momentum in the final states.  
The backgrounds of this process have much smaller rates than the signal one when the invariant mass of the $b\bar{b}$ pair ($M_{b\bar{b}}$) is around the $Z$-pole.  As mentioned earlier, the VBF$\to h \to b\bar{b}$ process can also be efficiently removed with an invariant mass window cut.  However, the events with high $b\bar{b}$ invariant mass values are also crucial for our analysis, for which the signal rates are much smaller, and a more careful treatment on the background is needed.  One potential background process is $\mu^- \mu^+ \to b \bar{b}$ with additional initial-state radiation (ISR) photons, which could fake the signal if the ISR photons are not tagged.  This background is highly suppressed when $M_{b\bar{b}}$ is much smaller than the center-of-mass energy of the collision.  As shown later in \autoref{sec:ana}, we will impose an upper bound on $M_{b\bar{b}}$ of 1\,TeV, which we expect to be able to efficiently remove this background.  
Another potentially sizable background is $\mu^- \mu^+ \to b\bar{b} \mu^-\mu^+$ via $ZZ/Z\gamma/\gamma\gamma$ fusion with undetected forward muons.  Assuming forward muon taggers will be available (see {\it e.g.}~Refs.~\cite{Ruhdorfer:2023uea, Ruhdorfer:2024dgz, Li:2024joa} for more details), only the muons in the very forward regions are undetected, so this background process will have a very small missing transverse momentum ($\slashed{p}_T$).  On the other hand, the signal events generally have a sizable $\slashed{p}_T$.  This is illustrated in 
\autoref{fig:pt_dis_bb_10}, where we assume muons with $\eta>6$ are untagged.  As such, this background can be efficiently removed with a mild $\slashed{p}_T$ cut ($\sim 20$\,GeV).  
Another potential background is the QCD parton induced
di-jet backgrounds~\cite{BuarqueFranzosi:2021wrv, Han:2021kes}, which we also expect to be efficiently removed with this $\slashed{p}_T$ cut.  
For convenience, we omit these background processes in our analysis and leave a more detailed background analysis to future studies.

\begin{figure}[t]
\centering

\subfigure[]{
\begin{minipage}[b]{0.3\textwidth}
\centering
\begin{tikzpicture}
	\begin{feynman}[small]
		\vertex (m1) ;
		\vertex [above left=of m1] (a) {\(\mu^{-}\)};
		\vertex [below left=of m1] (b) {\(\mu^{+}\)}; 
		\vertex [right=of m1] (m2);
		\vertex [above right=0.5 cm of m2] (m3);
		\vertex [below right=of m2] (c) {\(\nu_{\mu}\)};
		\vertex [right=of m3] (m4) [dot];
		\vertex [above right=0.5 cm of m3] (d) {\(\bar{\nu}_{\mu}\)};
		\vertex [above right=0.5 cm of m4] (e) {\(b\)};
		\vertex [below right=0.5 cm of m4] (f) {\(\bar{b}\)};
		
		\diagram* {
			(a) -- [fermion] (m1) -- [fermion] (b),
			(m1) -- [boson, edge label'=\(Z\)] (m2) -- [anti fermion, edge label'=\(\bar{\nu}_{\mu}\)] (m3) -- [boson, edge label=$Z$] (m4) -- [fermion] (e),
			(m2) -- [fermion] (c),
			(m3) -- [anti fermion] (d),
			(m4) [dot] -- [anti fermion] (f),
		};
	\end{feynman}
    \fill [black] (m4) circle (2pt);
\end{tikzpicture}
\end{minipage}}
\hfill
\subfigure[]{
\begin{minipage}[b]{0.3\textwidth}
\centering
\begin{tikzpicture}
	\begin{feynman}[small]
		\vertex (m1) ;
		\vertex [above left=of m1] (a) {\(\mu^{-}\)};
		\vertex [right=0.75of m1] (m2); 
		\vertex [above right=0.5 cm of m2] (b) {$b$};
		\vertex [below right=0.5 cm of m2] (c) {\(\bar{b}\)};
		\vertex [below=of m1] (m3);
		\vertex [below right=of m3] (d) {\(\nu_\mu\)};
		\vertex [below=of m3] (m4);
		\vertex [below left=of m4] (e) {\(\mu^+\)};
		\vertex [below right=of m4] (f) {\(\bar{\nu}_\mu\)};
		
		\diagram* {
			(a) -- [fermion] (m1) -- [boson, edge label=$Z/\gamma$] (m2),
			(b) -- [anti fermion] (m2) -- [anti fermion] (c),
			(m1) -- [fermion, edge label'=\(\mu^-\)] (m3) -- [boson, edge label'=\(W^{+}\)] (m4),
			(m3) -- [fermion] (d),
			(e) -- [anti fermion] (m4) -- [anti fermion] (f),
		};
	\end{feynman}
    \fill [black] (m2) circle (2pt);
\end{tikzpicture}
\end{minipage}}
\hfill
\subfigure[]{
\begin{minipage}[b]{0.3\textwidth}
\centering
\begin{tikzpicture}
	\begin{feynman}[small]
		\vertex (m1) ;
		\vertex [above left=of m1] (a) {\(\mu^{+}\)};
		\vertex [above right=of m1] (b) {\(\bar{\nu}_\mu\)}; 
		\vertex [below=of m1] (m2);
		\vertex [below left=of m2] (c) {\(\mu^{-}\)};	
		\vertex [right=of m2] (m3); 
		\vertex [right=3.5 cm of c] (d) {\(\nu_\mu\)};
		\vertex [above=of $(m3)!0.5!(d)$] (m4);
		\vertex [right=of m4, above=2cm of d] (e) {\(b\)};
		\vertex [right=of m4, above=of d] (f) {\(\bar{b}\)};
		
		\diagram* {
			(a) -- [anti fermion] (m1) -- [anti fermion] (b),
			(m1) -- [boson, edge label'=\(W^{+}\)] (m2),
			(c) -- [fermion] (m2) -- [fermion, edge label=\(\nu_\mu\)] (m3) -- [fermion] (d),
			(m3) -- [boson, edge label=\(Z\)] (m4),
			(m4) -- [fermion] (e),
			(m4) -- [anti fermion] (f),
		};
	\end{feynman}
    \fill [black] (m4) circle (2pt);
\end{tikzpicture}
\end{minipage}}

\subfigure[]{
\begin{minipage}[b]{0.3\textwidth}
\centering
\begin{tikzpicture}
	\begin{feynman}[small]
		\vertex (m1) ;
		\vertex [above left=of m1] (a) {\(\mu^{-}\)};
		\vertex [right=of m1] (m2); 
		\vertex at ($(a)!0.5!(m1)+ (1.5cm, 0)$) (m2);
		\vertex [right=3 cm of a] (b) {$b$};
		\vertex [right=of m1, below=of b] (c) {\(\bar{b}\)};
		\vertex [below=of m1] (m3);
		\vertex [below left=of m3] (d) {\(\mu^{+}\)};	
		\vertex [right=of m3] (m4); 
		\vertex at ($(d)!0.5!(m3)+ (1.5cm, 0)$) (m4);
		\vertex [right=3 cm of d] (f) {\(\bar{\nu}_\mu\)};
		\vertex [right=of m3, above=of f] (e) {$\nu_\mu$};
		
		\diagram* {
			(a) -- [fermion] (m1) -- [boson, edge label=$Z/\gamma$] (m2),
			(b) -- [anti fermion] (m2) -- [anti fermion] (c),
			(m1) -- [fermion, edge label'=$\mu^{+}$] (m3),
			(d) -- [anti fermion] (m3) -- [boson, edge label'=$Z$] (m4),
			(e) -- [anti fermion] (m4) -- [anti fermion] (f),
		};
	\end{feynman}
    \fill [black] (m2) circle (2pt);
\end{tikzpicture}
\end{minipage}}
\hfill
\subfigure[]{
\begin{minipage}[b]{0.3\textwidth}
\centering
\begin{tikzpicture}
    \begin{feynman}[small]
    	\vertex (m1) ;
    	\vertex [above left=of m1] (a) {\(\mu^{-}\)};
    	\vertex [above right=of m1] (b) {\(\nu_{\mu}\)}; 
    	\vertex [below=0.66 of m1] (m2);
    	\vertex [below=0.66 of m2] (m3);
    	\vertex [below=0.66 of m3] (m4);
    	\vertex [below right=of m4] (c) {\(\bar{\nu}_{\mu}\)};
    	\vertex [below left=of m4] (d) {\(\mu^{+}\)};
    	\vertex [right=of m2] (e) {\(b\)};
    	\vertex [right=of m3] (f) {\(\bar{b}\)};
    	
    	\diagram* {
    		(a) -- [fermion] (m1) -- [fermion] (b),
    		(m1) -- [boson, edge label'=\(W^{-}\)] (m2) -- [anti fermion, edge label=\(t\)] (m3) -- [boson, edge label'=\(W^{+}\)] (m4),
    		(c) -- [fermion] (m4) -- [fermion] (d),
    		(m2) -- [fermion] (e),
    		(m3) -- [anti fermion] (f),
    	};
    \end{feynman}
    \fill [black] (m2) circle (2pt);
    \fill [black] (m3) circle (2pt);
\end{tikzpicture}
\end{minipage}}
\hfill
\subfigure[]{
\begin{minipage}[b]{0.3\textwidth}
\centering
\begin{tikzpicture}
	\begin{feynman}[small]
		\vertex (m1) ;
		\vertex [above left=of m1] (a) {\(\mu^{-}\)};
		\vertex [above right=of m1] (b) {\(\nu_{\mu}\)}; 
		\vertex [below=of m1] (m2);
		\vertex [below=of m2] (m3);
		\vertex [below right=of m3] (c) {\(\bar{\nu}_{\mu}\)};
		\vertex [below left=of m3] (d) {\(\mu^{+}\)};
		\vertex [right=.75 of m2] (m4);
		\vertex [below right=0.5 cm of m4] (e) {\(\bar{b}\)};
		\vertex [above right=0.5 cm of m4] (f) {\(b  \)};
		
		\diagram* {
			(a) -- [fermion] (m1) -- [fermion] (b),
			(m1) -- [boson, edge label'=\(W^{-}\)] (m2) -- [boson, edge label'=\(W^{+}\)] (m3),
			(c) -- [anti fermion] (m3) -- [anti fermion] (d),
			(m2) -- [boson, edge label=\(Z/\gamma\)] (m4),
			(e) -- [anti fermion] (m4) -- [anti fermion] (f),
		};
	\end{feynman}
    \fill [black] (m4) circle (2pt);
\end{tikzpicture}
\end{minipage}}

\caption{Typical Feynman diagrams for the process $\mu^{-}\mu^{+}\rightarrow b\bar{b}\nu_{\mu}\bar{\nu}_{\mu}$. Diagram (a) to (c) are examples of $2f\to 2f$ processes with an additional initial or final state $Z/\gamma$. Other diagrams of the same type are not explicitly shown.  
Diagram (d) is neutral diboson production, while (e) and (f) is VBF. 
BSM vertexes are indicated by black dots.
}
\label{fig:FDbb}
\end{figure}

\begin{figure}[htb]
    \centering
	\includegraphics[width=0.70\textwidth]{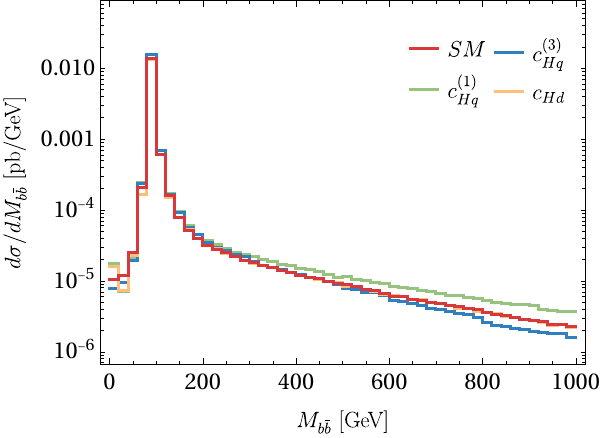}
\caption{The differential Cross Section distribution on the invariant mass of $b\bar{b}$ pair ($M_{b\bar{b}}$) for the $\mu^-\mu^+ \rightarrow b\bar{b}\nu_{\mu}\bar{\nu}_{\mu}$ process with  $\sqrt{s}=10\,\TeV$.  A bin width of 20\,GeV is chosen.  The red curve corresponds to the SM value while the other three curves each corresponds to one of the three Wilson coefficients $c_{H q}^{(1)} ,\, c_{H q}^{(3)} ,\,  c_{H d}$ setting to a reference value of 1 (with $\Lambda=1\,\TeV$) while the other two are set to zero.  
Note that the low energy bins (with $M_{b\bar{b}}\ll m_Z$) are subject to large statistical uncertainties in our simulation.
}
\label{fig:inv_dis_bb_10}
\end{figure}

The differential distribution of the $\mu^-\mu^+ \to b\bar{b}\nu_{\mu}\bar{\nu}_{\mu}$ process contains crucial information. 
In \autoref{fig:inv_dis_bb_10} we show the binned differential cross section as a function of the invariant mass of the $b\bar{b}$ pair. The red curve corresponds to the SM value while the other three curves each corresponds to one of the three Wilson coefficients $c_{H q}^{(1)} ,\, c_{H q}^{(3)} ,\,  c_{H d}$ 
setting to a reference value of 
1 (with $\Lambda=1\,$TeV) while the other two are set to zero. 
A peak around the Z-pole can be clearly seen, which corresponds to the $s$-channel $Z$ diagram in \autoref{fig:FDbb}(f).  The cross section also drops rapidly as the invariant mass increases -- a typical feature for the $WW$ fusion processes.   
Nevertheless, the events with high invariant mass contain important information that is complementary to the ones around the Z-pole.  In particular, as shown in \autoref{fig:inv_dis_bb_10}, the cross section becomes more sensitive to the coefficients $c_{H q}^{(1)}$ and $c_{H q}^{(3)}$ at higher invariant mass, a feature expected from their energy enhancements.  As we will show later, a binned analysis that takes account of the information in the invariant mass distribution makes it possible to simultaneously constrain all three Wilson coefficients.

\begin{figure}[htb]
\centering
\subfigure{
    \includegraphics[width=0.30\textwidth]{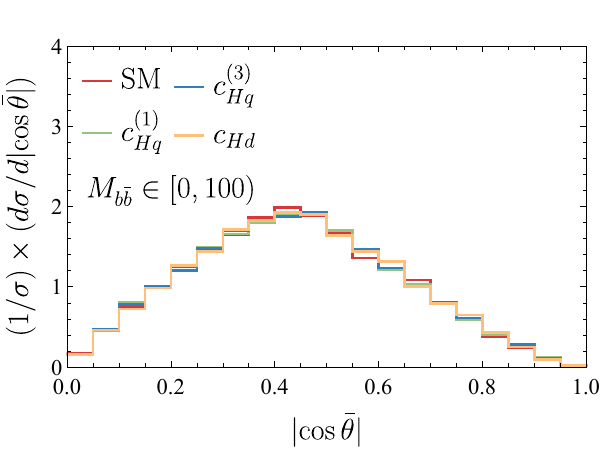}
    \label{subfig:chq1chd_10TeV}
}
\subfigure{
    \includegraphics[width=0.30\textwidth]{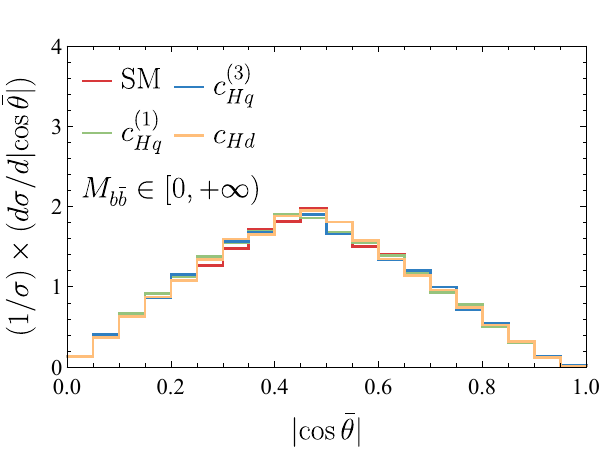}
    \label{subfig:chq3chd_10TeV}
}
\subfigure{
    \includegraphics[width=0.30\textwidth]{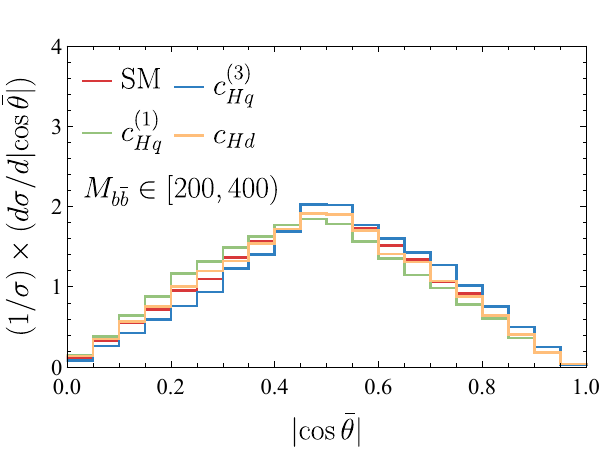}
    \label{subfig:chq1chq3_30TeV}
}
\subfigure{
    \includegraphics[width=0.30\textwidth]{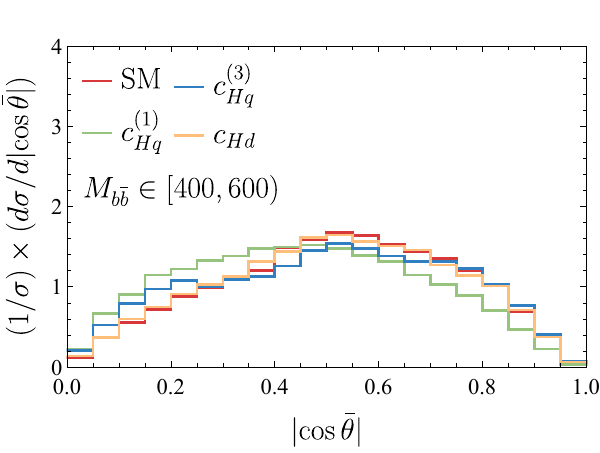}
    \label{subfig:chq1chd_30TeV}
}
\subfigure{
    \includegraphics[width=0.30\textwidth]{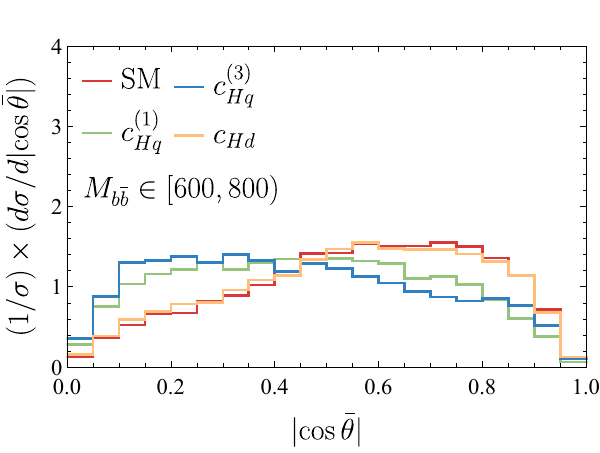}
}
\subfigure{
    \includegraphics[width=0.30\textwidth]{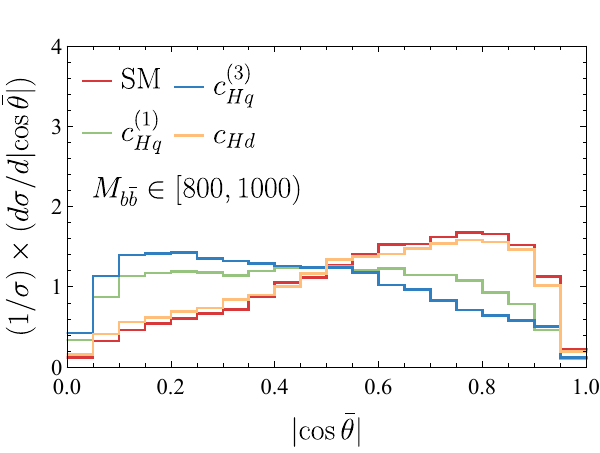}
}
\subfigure{
    \includegraphics[width=0.30\textwidth]{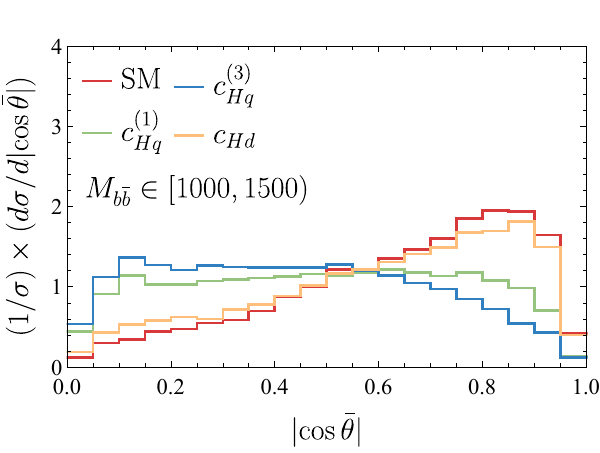}
}
\subfigure{
    \includegraphics[width=0.30\textwidth]{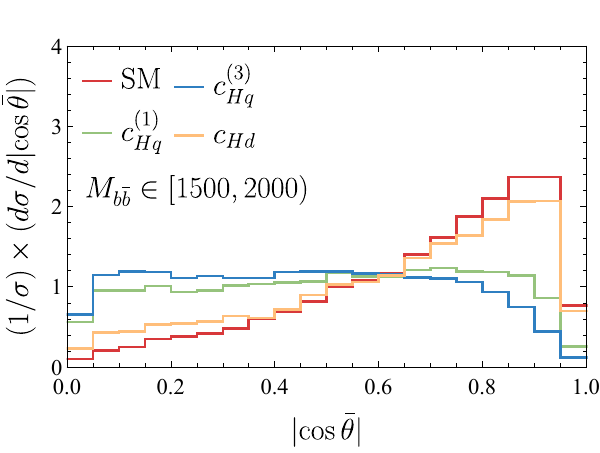}
}
\subfigure{
    \includegraphics[width=0.30\textwidth]{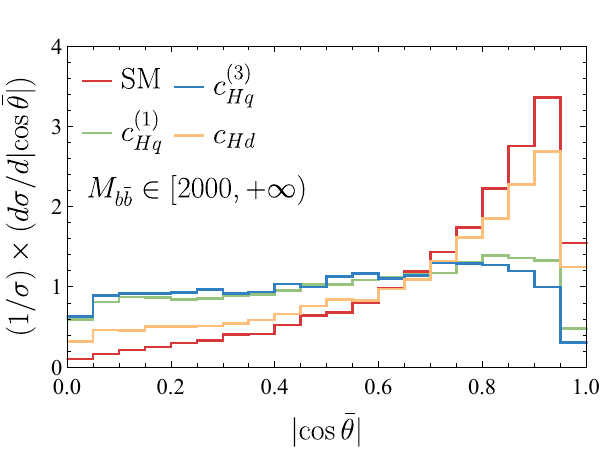}
}
\caption{Differential cross section of the $\cosbt$ variable, defined as $\cosbt\equiv \frac{1}{2}\left(\left|\cos\theta_{\mu^{-}b}\right|+\left|\cos\theta_{\mu^{+}b}\right|\right)$, where $\theta_{\mu^-b}$ ($\theta_{\mu^+b}$) is the angle between $\mu^-$ ($\mu^+$) and $b$ in the c.o.m.\ frame of the $b\bar{b}$ system.  
A reference value of 10 is chosen here for each Wilson coefficient (assuming $\Lambda=1\,$TeV, and setting the other two to zero), with only linear contribution considered.  The bin width is chosen to be 0.05.  A (different) invariant mass selection is applied for each of the 9 plots, which corresponds to the 9 bins listed in \autoref{tab:invbin}. 
}
\label{fig:cos_dis_bb10}
\end{figure}

The angular distributions of the $\mu^-\mu^+ \to b\bar{b}\nu_{\mu}\bar{\nu}_{\mu}$ process also contain useful information. For instance, the diagram with the $s$-channel $Z$ would have different kinematics with the one with a $t$-channel fermion exchange.  Ideally, this information is captured by the angle between the incoming $W^\pm$ and the outgoing $b$ or $\bar{b}$. However, this angle cannot be directly measured due to the missing neutrinos. Nevertheless, part of the information is still in the production polar angles of $b$ and $\bar{b}$. In \autoref{fig:cos_dis_bb10}, we show the distribution of the variable $\cosbt$ for events in different ranges of $M_{b\bar{b}}$  (same as the choices of 9 bins listed in \autoref{tab:invbin} of the following Section), where $\theta$ is the polar angle of $b$ in the center-of-mass frame of the $b\bar{b}$ system.  The reason for going to the $b\bar{b}$ c.o.m.\ frame is that the kinematics in the lab frame is subject to a large longitudinal boost, similar to the situation of LHC.  Note that, in the $b\bar{b}$ c.o.m.\ frame, the incoming $\mu^-$ and $\mu^+$ are in general not on the same line, and $\cosbt$ is simply defined as the average of the two different values, 
\beq
\cosbt\equiv \frac{1}{2}\left(\left|\cos\theta_{\mu^{-}b}\right|+\left|\cos\theta_{\mu^{+}b}\right|\right) \,, \label{eq:cosbartheta}
\eeq
where $\theta_{\mu^-b}$ ($\theta_{\mu^+b}$) is the angle between $\mu^-$ ($\mu^+$) and $b$.   Here, for the three Wilson coefficients $c_{H q}^{(1)} ,\, c_{H q}^{(3)} ,\,  c_{H d}$, a reference value of 10 is chosen instead (assuming $\Lambda=1\,$TeV, and in each case the other two are set to zero) to better visualize their effects.\footnote{We also consider only linear contributions of the Wilson coefficients.  Note that in the analysis the typical constraints on the Wilson coefficients are much smaller than 10, and their quadratic contributions are negligible.} 
It is clear in \autoref{fig:cos_dis_bb10} that, for large $M_{b\bar{b}}$ (much above $Z$ pole), $\cosbt$ is sensitive to $c_{H q}^{(1)}$ and $c_{H q}^{(3)}$, which give a more even distribution, while the SM one concentrates more in the forward region (since the SM cross section is dominated by the $t$-channel diagram at high $M_{b\bar{b}}$).  As such, for large $M_{b\bar{b}}$ we can further use the $\cosbt$ distribution to increase the sensitivity to the Wilson coefficients.

The discussion above also applies to the $\mu^-\mu^+ \to c\bar{c}\nu_{\mu}\bar{\nu}_{\mu}$ and $\mu^-\mu^+ \to \tau^{-}\tau^{+}\nu_{\mu}\bar{\nu}_{\mu}$, which have very similar features.  For the asymmetric processes $\mu^-\mu^+ \to c s \nu_{\mu} \mu$ and $\mu^-\mu^+ \to \tau \nu_\tau \nu_{\mu} \mu$, the situations are slightly different.  Taking $\tau \nu_\tau \nu_{\mu} \mu$ for example, the typical diagrams are shown in \autoref{fig:FDmuta}, where again the dominant contribution to the total cross section comes from the $WZ/W\gamma$ fusion diagrams (f), (g).\footnote{Note that, while the diboson process in \autoref{fig:FDmuta} (d) and (e) is very sensitive to a number of operators, in our analysis the only SMEFT contribution is the modification of the $W\tau\nu_\tau$ coupling, in which case the diboson process plays a less important row due to its smaller cross section.  Once again, all diagrams to the same $4f$ final states are included in our analysis.}  They are also more sensitive to the $\mathcal{O}^{(3)}_{Hl}$ ($\mathcal{O}^{(3)}_{Hq}$ for $c s \nu_{\mu} \mu$) operator which modifies the $W$-fermion couplings. Again, we do not expect any significant background for this process.  
While the invariant mass distribution of the $\tau\nu_\tau$ pair still contains useful information and peaks around the $W$ mass, it cannot be reconstructed due to the additional missing $\nu_\mu$.  Nevertheless, the measurement of the total rate of this process contains useful information which is complementary to the one of $\mu^-\mu^+ \to \tau^{-}\tau^{+}\nu_{\mu}\bar{\nu}_{\mu}$, as we will show later. 
Similarly, the measurement of $cs\nu_\mu\mu$ process also provides information complementary to the one of $c \bar{c}\nu_{\mu}\bar{\nu}_{\mu}$.

\begin{figure}[t]
\centering

\subfigure[]{
\begin{minipage}[b]{0.3\textwidth}
\centering
\begin{tikzpicture}
	\begin{feynman}[small]
		\vertex (m1) ;
		\vertex [above left=of m1] (a) {\(\mu^{+}\)};
		\vertex [below left=of m1] (b) {\(\mu^{-}\)}; 
		\vertex [right=of m1] (m2);
		\vertex [above right=0.5 cm of m2] (m3);
		\vertex [below right=of m2] (c) {\(\bar{\nu}_{\mu}\)};
		\vertex [right=of m3] (m4) [dot];
		\vertex [above right=0.5 cm of m3] (d) {\(\mu^{-}\)};
		\vertex [above right=0.5 cm of m4] (e) {\(\tau^{+}\)};
		\vertex [below right=0.5 cm of m4] (f) {\(\nu_{\tau}\)};
		
		\diagram* {
			(a) -- [fermion] (m1) -- [fermion] (b),
			(m1) -- [boson, edge label'=\(Z\)] (m2) -- [fermion, edge label=\(\nu_{\mu}\)] (m3) -- [boson, edge label'=\(W^{+}\)] (m4) -- [anti fermion] (e),
			(m2) -- [anti fermion] (c),
			(m3) -- [fermion] (d),
			(m4) [dot] -- [fermion] (f),
		};
	\end{feynman}
    \fill [black] (m4) circle (2pt);
\end{tikzpicture}
\end{minipage}}
\hfill
\subfigure[]{
\begin{minipage}[b]{0.3\textwidth}
\centering
\begin{tikzpicture}
	\begin{feynman}[small]
		\vertex (m1) ;
		\vertex [above left=of m1] (a) {\(\mu^{+}\)};
		\vertex [right=0.75of m1] (m2); 
		\vertex [above right=0.5 cm of m2] (b) {\(\tau^{+}\)};
		\vertex [below right=0.5 cm of m2] (c) {\(\nu_{\tau}\)};
		\vertex [below=of m1] (m3);
		\vertex [right=of m3] (d) {\(\bar{\nu}_\mu\)};
		\vertex [below=of m3] (m4);
		\vertex [below left=of m4] (e) {\(\mu^-\)};
		\vertex [below right=of m4] (f) {\(\mu^{-}\)};
		
		\diagram* {
			(a) -- [anti fermion] (m1) -- [boson, edge label=$W^{+}$] (m2),
			(b) -- [fermion] (m2) -- [fermion] (c),
			(m1) -- [anti fermion, edge label'=\(\nu_{\mu}\)] (m3) -- [boson, edge label'=\(Z\)] (m4),
			(m3) -- [anti fermion] (d),
			(e) -- [fermion] (m4) -- [fermion] (f),
		};
	\end{feynman}
    \fill [black] (m2) circle (2pt);
\end{tikzpicture}
\end{minipage}}
\hfill
\subfigure[]{
\begin{minipage}[b]{0.3\textwidth}
\centering
\begin{tikzpicture}
	\begin{feynman}[small]
		\vertex (m1) ;
		\vertex [above left=of m1] (a) {\(\mu^{+}\)};
		\vertex [above right=of m1] (b) {\(\bar{\nu}_\mu\)}; 
		\vertex [below=of m1] (m2);
		\vertex [below left=of m2] (c) {\(\mu^{-}\)};	
		\vertex [right=of m2] (m3); 
		\vertex [right=3.5 cm of c] (d) {\(\mu^{-}\)};
		\vertex [above=of $(m3)!0.5!(d)$] (m4);
		\vertex [above right=0.5 of m4] (e) {\(\tau^{+}\)};
		\vertex [below right=0.5 of m4] (f) {\(\nu_{\tau}\)};
		
		\diagram* {
			(a) -- [anti fermion] (m1) -- [anti fermion] (b),
			(m1) -- [boson, edge label'=\(W^{-}\)] (m2),
			(c) -- [fermion] (m2) -- [fermion, edge label'=\(\nu_{\mu}\)] (m3) -- [fermion] (d),
			(m3) -- [boson, edge label=\(W^{+}\)] (m4),
			(m4) -- [anti fermion] (e),
			(m4) -- [fermion] (f),
		};
	\end{feynman}
    \fill [black] (m4) circle (2pt);
\end{tikzpicture}
\end{minipage}}

\subfigure[]{
\begin{minipage}[b]{0.22\textwidth}
\centering
\begin{tikzpicture}
	\begin{feynman}[small]
		\vertex (m1) ;
		\vertex [above left=of m1] (a) {\(\mu^{+}\)};
		\vertex [right=of m1] (m2); 
		\vertex at ($(a)!0.5!(m1)+ (1.5cm, 0)$) (m2);
		\vertex [right=3 cm of a] (b) {$\tau^{+}$};
		\vertex [right=of m1, below=of b] (c) {\(\nu_{\tau}\)};
		\vertex [below=of m1] (m3);
		\vertex [below left=of m3] (d) {\(\mu^{-}\)};	
		\vertex [right=of m3] (m4); 
		\vertex at ($(d)!0.5!(m3)+ (1.5cm, 0)$) (m4);
		\vertex [right=3 cm of d] (f) {\(\bar{\nu}_{\mu}\)};
		\vertex [right=of m3, above=of f] (e) {$\mu^{-}$};
		
		\diagram* {
			(a) -- [anti fermion] (m1) -- [boson, edge label=$W^{+}$] (m2),
			(b) -- [fermion] (m2) -- [fermion] (c),
			(m1) -- [anti fermion, edge label'=$\nu_{\mu}$] (m3),
			(d) -- [fermion] (m3) -- [boson, edge label'=$W^{-}$] (m4),
			(e) -- [anti fermion] (m4) -- [anti fermion] (f),
		};
	\end{feynman}
    \fill [black] (m2) circle (2pt);
\end{tikzpicture}
\end{minipage}}
\hfill
\subfigure[]{
\begin{minipage}[b]{0.24\textwidth}
\centering
\begin{tikzpicture}
	\begin{feynman}[small]
		\vertex (m1) ;
		\vertex [above left=of m1] (a) {\(\mu^{+}\)};
		\vertex [below left=of m1] (b) {\(\mu^{-}\)}; 
        \vertex [right=of m1] (m2);
		\vertex [above right=of m2] (m3);
		\vertex [below right=of m2] (m4);
        \vertex at ($(m3) + (1cm, 0.5cm)$) (c) {\(\tau^+\)};
		\vertex at ($(m3) + (1cm, -0.5cm)$) (d) {\(\nu_\tau\)};
        \vertex at ($(m4) + (1cm, 0.5cm)$) (e) {\(\bar{\nu}_{\mu}\)};
        \vertex at ($(m4) + (1cm, -0.5cm)$) (f) {\(\mu^{-}\)};
		
		\diagram* {
			(a) -- [anti fermion] (m1) -- [anti fermion] (b),
            (m1) -- [photon, edge label=\(Z/\gamma\)] (m2),
            (m2) -- [photon, edge label=\(W^{+}\)] (m3),
            (m2) -- [photon, edge label'=\(W^{-}\)] (m4),
            (c) -- [fermion] (m3) -- [fermion] (d),
            (e) -- [fermion] (m4) -- [fermion] (f),
		};
	\end{feynman}
    \fill [black] (m3) circle (2pt);
\end{tikzpicture}
\end{minipage}}
\hfill
\subfigure[]{
\begin{minipage}[b]{0.22\textwidth}
\centering
\begin{tikzpicture}
    \begin{feynman}[small]
    	\vertex (m1) ;
    	\vertex [above left=of m1] (a) {\(\mu^{+}\)};
    	\vertex [above right=of m1] (b) {\(\bar{\nu}_{\mu}\)}; 
    	\vertex [below=0.66 of m1] (m2);
    	\vertex [below=0.66 of m2] (m3);
    	\vertex [below=0.66 of m3] (m4);
    	\vertex [below right=of m4] (c) {\(\mu^{-}\)};
    	\vertex [below left=of m4] (d) {\(\mu^{-}\)};
    	\vertex [right=of m2] (e) {\(\tau^{+}\)};
    	\vertex [right=of m3] (f) {\(\nu_{\tau}\)};
    	
    	\diagram* {
    		(a) -- [anti fermion] (m1) -- [anti fermion] (b),
    		(m1) -- [boson, edge label'=\(W^{+}\)] (m2) -- [fermion, edge label=\(\bar{\nu}_{\tau}\)] (m3) -- [boson, edge label'=\(Z\)] (m4),
    		(c) -- [anti fermion] (m4) -- [anti fermion] (d),
    		(m2) -- [anti fermion] (e),
    		(m3) -- [fermion] (f),
    	};
    \end{feynman}
    \fill [black] (m2) circle (2pt);
    \fill [black] (m3) circle (2pt);
\end{tikzpicture}
\end{minipage}}
\hfill
\subfigure[]{
\begin{minipage}[b]{0.22\textwidth}
\centering
\begin{tikzpicture}
	\begin{feynman}[small]
		\vertex (m1) ;
		\vertex [above left=of m1] (a) {\(\mu^{+}\)};
		\vertex [above right=of m1] (b) {\(\bar{\nu}_{\mu}\)}; 
		\vertex [below=of m1] (m2);
		\vertex [below=of m2] (m3);
		\vertex [below right=of m3] (c) {\(\mu^{-}\)};
		\vertex [below left=of m3] (d) {\(\mu^{-}\)};
		\vertex [right=.75 of m2] (m4);
		\vertex [below right=0.5 cm of m4] (e) {\(\nu_{\tau}\)};
		\vertex [above right=0.5 cm of m4] (f) {\(\tau^{+}\)};
		
		\diagram* {
			(a) -- [anti fermion] (m1) -- [anti fermion] (b),
			(m1) -- [boson, edge label'=\(W^{+}\)] (m2) -- [boson, edge label'=\(Z/\gamma\)] (m3),
			(c) -- [anti fermion] (m3) -- [anti fermion] (d),
			(m2) -- [boson, edge label=\(W^{+}\)] (m4),
			(e) -- [anti fermion] (m4) -- [anti fermion] (f),
		};
	\end{feynman}
    \fill [black] (m4) circle (2pt);
\end{tikzpicture}
\end{minipage}}

\caption{Typical Feynman diagram for the process $\mu^{-}\mu^{+} \rightarrow \tau^+ \nu_\tau \bar{\nu}_{\mu} \mu^- $. Diagram (a) to (c) are examples of $2f\to 2f$ processes with additional $W$ in the initial or final state, 
diagram (d) and (e) are diboson production, while (f) and (g) is VBF. 
BSM vertexes are indicated by black dots.
}
\label{fig:FDmuta}
\end{figure}

\section{Measurements and Analyses}
\label{sec:ana}

\subsection{Run scenarios}

We consider two runs with center-of-mass energy 10\,TeV and 30\,TeV.  
The reference values of the integrated luminosity are taken from the ideal luminosity relations~\cite{Accettura:2023ked},

\beq
\mathcal{L}_{int} = 10 \ab^{-1} \left( \frac{E_{cm}}{10\TeV} \right)^2 \,,
\label{eq:lumi}
\eeq 
which are $10\inab$ for the 10\,TeV run and $90\inab$ for the 30\,TeV run, respectively.  
It is worth noting that the run scenarios considered in this work are optimistic, and are subject to change with the continuous development of the muon collider R\&D program.

\subsection{Methodology}
\label{subsec:method}

We simulate our signal processes with \textsc{Mad}\textsc{Graph5}\_\textsc{aMC}@NLO v3.5.6~\cite{Alwall:2014hca} at parton level with SMEFT effects implemented by SMEFTsim 3.0 package~\cite{Brivio:2020onw} as a UFO model~\cite{Darme:2023jdn}. We apply the following cuts within this study: 
$p^j_T>20$\,GeV, $|\eta_j|<2.44$ for jets and $p^l_T>10$\,GeV, $|\eta_\tau|<2.44$, $|\eta_\mu|<6$ for charged leptons~\cite{Han:2021kes,BuarqueFranzosi:2021wrv,Li:2024joa}. 
The choices of rapidity cuts are typical for the analyses at muon colliders, since it is difficult to detect jets and charged leptons (except muon) close to the beam due to the dramatic beam induced background (BIB) \cite{Li:2024joa}.  We also assume that forward muon taggers~\cite{Ruhdorfer:2023uea, Ruhdorfer:2024dgz, Li:2024joa} are implemented, which results in the much larger acceptance for $\eta_\mu$.  
We do not expect these cuts to have very significant impacts in our analysis except for the asymmetric processes $c s \nu_{\mu} \mu$ and $\tau \nu_\tau \nu_{\mu} \mu$, for which we require the muon to be tagged.  As such, the selection efficiency of the asymmetric processes are very sensitive to the performance of the forward muon taggers. 
We do not expect the detector effects to have a significant impact in our analysis either, as we use only simple kinematic variables including the invariant mass and the $\cosbt$ variable defined in \autoref{eq:cosbartheta}.  However, the tagging efficiencies of $b$, $c$, $\tau$ are crucial in our analysis since they have a directly impact on the total rate.  Ref.~\cite{Accettura:2023ked} provides a benchmark efficiency of around 80\% for tagging one $b$ jet, and around 70\% for the di-tau system.  For the charm and strange quarks, the tagging efficiencies are not listed in Ref.~\cite{Accettura:2023ked}, and we use the recent CEPC study~\cite{Liang:2023wpt} as a reference (assuming a muon collider could achieve similar performances), which are approximately $70\%$ for a charm jet and $50\%$ for a strange jet, respectively.\footnote{For the charm-tagging rate, Refs.~\cite{CLICdp:2018cto,Suehara:2015ura} suggest a reference value of around 67\% for CLIC, which is close to 70\%. } 
Note that, these tagging rates could be somewhat optimistic for muon colliders, and should be replaced by more realistic estimations in future studies once they become available.  
Apart from the $\tau \nu_\tau \nu_{\mu} \mu$ process, we always require the two final state fermions to be both tagged, which gives {\it e.g.} a $64\%$ efficiency for the $b\bar{b}$ pair. For $\tau \nu_\tau \nu_{\mu} \mu$, we assume that the final states can be selected by requiring one $\tau$-tag in addition to a muon and missing momentum, and naively assign an efficiency of $\sqrt{0.7}\approx0.84$.  We do not require the discrimination between $b$ and $\bar{b}$ (or $c$ and $\bar{c}$) since this information is not used in our analysis.  The signal selection efficiencies based on tagging are summarized in \autoref{tab:eff}.
We also do not expect mis-tagging to have a significant impact on our analyses, given that we do not have any overwhelming backgrounds to start with.  The only sizable mis-tagging rate is the one for $b$ to be identified as $c$, which is around 20\%~\cite{Accettura:2023ked}.  Given that we require both final state quarks to be tagged, we only expect a small mixing ($\lesssim 4\%$) between different signal events, which we simply ignore in our study.  Once all the mis-tagging rates are available, one could also implement their effects on the signal events (which becomes a mixture of different processes) under a global SMEFT framework, which is beyond the scope of our current analysis.    
\begin{table}[htb]
\centering
\begin{tabular}{|c|c|c|}\hline
Process & Requirement & Efficiency \\ \hline \hline
$\mu^-\mu^+ \to b \bar{b}\nu_{\mu}\bar{\nu}_{\mu}$ & 2 b-tags & 0.64 \\ \hline
$\mu^-\mu^+ \to c \bar{c}\nu_{\mu}\bar{\nu}_{\mu}$ & 2 c-tags & 0.49\\ \hline
$\mu^-\mu^+ \to \tau^- \tau^+\nu_{\mu}\bar{\nu}_{\mu}$ & 2 $\tau$-tags & 0.7\\ \hline
$\mu^-\mu^+ \to c s \nu_{\mu} \mu$ & 1 c-tag and 1 s-tag & 0.35\\ \hline
$\mu^-\mu^+ \to \tau \nu_\tau \nu_{\mu} \mu$ & 1 $\tau$-tag & 0.84\\ \hline
\end{tabular}
\caption{A summary of the tagging efficiencies of different processes implemented in our analysis.  }
\label{tab:eff}
\end{table}

We consider the leading SMEFT contributions at the $\Lambda^{-2}$ order.  At the cross section level, this means that only the linear contributions of the Wilson coefficients are considered, while the quadratic contributions are omitted since they are at the $\Lambda^{-4}$ order.  Given the high measurement precision, this is a very good approximation in our analysis.  The SMEFT prediction of the cross sections are thus parameterized as
\beq
\sigma_{\rm SMEFT} = \sigma_{\rm SM} + \sum_i \alpha_i c_i + \mathcal{O}(\Lambda^{-4}) \,, \label{eq:sigmac}
\eeq
where the coefficients $\alpha_i$ can be determined numerically from the MC simulation.  The full list of the numerical expressions of \autoref{eq:sigmac} for all the processes and bins are provided in \ref{app:xs}.   
To extract the bounds on the Wilson coefficients, we implement the chi-squared method, with the total $\chi^2$ given by
\begin{equation}\label{eq:chisq}
	\chi^2 = \sum_i^{\rm bins} \frac{(\sigma_{i,\rm SMEFT} - \sigma_{i,\rm exp})^2}{(\Delta\sigma_i)^2} \,,
\end{equation}
where the sum is performed over all the bins (to be specified later), and $\sigma_{i,\rm SMEFT}$, $\sigma_{i,\rm exp}$ and $\Delta\sigma_i$ are the SMEFT prediction, the measured central value and the (one-sigma) uncertainty of the cross section in the $i$th bin, respectively. 
 By construction, we assume the measured central values are SM-like, $\sigma_{i,\rm exp}=\sigma_{i,\rm SM}$.  
Note that we have also explicitly assumed that the measurements of different bins are uncorrelated.  We consider only statistical uncertainties, in which case
\beq
\Delta \sigma_i = \frac{\sigma_{i,\rm SM}}{\sqrt{N_{i,\rm SM}}} = \sqrt{\frac{\sigma_{i,\rm SM}}{L}} \,,
\eeq
where $N_{i,\rm SM}$ is the number of events in the bin and $L$ is the total luminosity.  

The inverse of covariant matrix of the Wilson coefficients can be estimated as
\begin{equation}
	(U^{-1})_{ij} = \frac{1}{2} \left.\frac{\partial^2 \chi^2}{\partial c_i \partial c_j}\right|_{\boldsymbol{c}=\hat{\boldsymbol{c}}} \,,
\end{equation}
where $c_i$ are the Wilson coefficients and $\hat{\boldsymbol{c}}$ their best fitted values (with minimum $\chi^2$), which are zero in our analysis by construction.  The one sigma bounds $\delta c_i$ and the correlation matrix $\rho_{ij}$ of the Wilson coefficients can be obtained as
\begin{equation}
	\delta c_i = \sqrt{U_{ii}} \,,
\end{equation}
and 
\begin{equation}
	\rho_{ij} = \frac{U_{ij}}{\delta c_i \delta c_j} \,.
\end{equation}

\subsection{Binning}
\label{subsec:bin}

For the symmetric processes ($\mu^-\mu^+ \to f \bar{f}\nu_{\mu}\bar{\nu}_{\mu}$ where $f \bar{f}=b\bar{b}$, $c\bar{c}$ or $\tau^-\tau^+$), the invariant mass distribution of the $f\bar{f}$ pair contain crucial important information in probing the Wilson coefficients.  To extract this information, we perform a binned analysis by dividing the signal process into 9 bins, with the range of the bins listed in \autoref{tab:invbin}.  The first bin has the largest cross section due to the $Z$ resonance.  To ensure enough MC statistics for higher invariant mass bins, each bin is separately generated in MadGraph with the corresponding invariant mass cuts.    

As mentioned earlier, for very high invariant mass, the signal could be subject to a sizable $\mu^-\mu^+ \to f\bar{f}$ background, and it is desirable to impose a upper bound on $M_{f\bar{f}}$.  Furthermore, the validity of the EFT analysis also tends to be problematic for high $M_{f\bar{f}}$, given that the typical reach on the new physics scale $\Lambda$ could be comparable or even smaller than $M_{f\bar{f}}$~\cite{Contino:2016jqw,Alte:2017pme}.  To resolve both issues, after we study the effects of the invariant mass bins in \autoref{subsec:invbin}, we will impose an upper bound of 1\,TeV on $M_{f\bar{f}}$ and discard bin 7-9 listed in \autoref{tab:invbin}.

\begin{table}
\resizebox{\linewidth}{!}{
\begin{tabular}{@{}m{3cm}<{\centering}|m{2cm}<{\centering}|m{2cm}<{\centering}|m{2cm}<{\centering}|m{2cm}<{\centering}|m{2cm}<{\centering}|m{2cm}<{\centering}|m{2cm}<{\centering}|m{2cm}<{\centering}|m{2cm}<{\centering}@{}}
\toprule
Bin number           & bin 1     & bin 2       & bin 3       & bin 4       & bin 5       & bin 6        & bin 7         & bin 8         & bin 9            \\ \midrule
Invariant mass & \multirow{2}{*}{$[0,100)$} & \multirow{2}{*}{$[100,200)$} & \multirow{2}{*}{$[200,400)$} & \multirow{2}{*}{$[400,600)$} & \multirow{2}{*}{$[600,800)$} & \multirow{2}{*}{$[800,1000)$} & \multirow{2}{*}{$[1000,1500)$} & \multirow{2}{*}{$[1500,2000)$} & \multirow{2}{*}{$[2000,+\infty)$} \\ 
(GeV) & & & & & & & & & \\ \bottomrule
\end{tabular}
}
\caption{The binning of the invariant mass of the $f\bar{f}$ pair for $\mu^-\mu^+ \to f \bar{f}\nu_{\mu}\bar{\nu}_{\mu}$, where $f \bar{f}=b\bar{b}$, $c\bar{c}$ or $\tau^-\tau^+$.
Note that for all results except those in \autoref{subsec:invbin}, 
we will impose the cut $M_{f\bar{f}}<1$\,TeV and discard bin 7-9.
}
\label{tab:invbin}
\end{table}

In addition, we will also make use of the $\cosbt$ variable in \autoref{eq:cosbartheta}.  To capture the essential information shown in \autoref{fig:cos_dis_bb10}, the events in each invariant mass bin starting from 600\,GeV are further divided into two bins, with the division point as listed in \autoref{tab:divide}.  The values of the division points are determined based a simple optimization procedure (by looking at a few benchmark values and choosing the one with the best result). 
Finally, the total $\chi^2$ is obtained by summing over the $\chi^2$ of all bins using \autoref{eq:chisq}.  
As shown in the next section, the binning is very helpful in discriminating the contributions of different Wilson coefficients and can significantly improve the overall results.   

\begin{table}[b]
\begin{tabular}{m{3cm}<{\centering}|cm{2cm}<{\centering}|cm{2cm}<{\centering}|cm{2cm}<{\centering}}
\toprule
{Invariant Mass} & \multicolumn{2}{c|}{$b\bar{b}$} & \multicolumn{2}{c|}{$c\bar{c}$} & \multicolumn{2}{c}{$\tau^{-}\tau^{+}$} \\
{(GeV)} & \multicolumn{1}{c}{\SI{10}{TeV}} & \SI{30}{TeV} & \multicolumn{1}{c}{\SI{10}{TeV}} & \SI{30}{TeV} & \multicolumn{1}{c}{\SI{10}{TeV}} & \SI{30}{TeV} \\ \midrule
{[}600, 800) & \multicolumn{1}{m{2cm}<{\centering}}{0.45} & 0.5 & \multicolumn{1}{m{2cm}<{\centering}}{0.45} & 0.45 & \multicolumn{1}{m{2cm}<{\centering}}{0.5} & 0.5 \\
{[}800, 1000) & \multicolumn{1}{c}{0.55} & 0.5 & \multicolumn{1}{c}{0.45} & 0.45 & \multicolumn{1}{c}{0.5} & 0.5 \\
{[}1000, 1500) & \multicolumn{1}{c}{0.6} & 0.5 & \multicolumn{1}{c}{0.45} & 0.5 & \multicolumn{1}{c}{0.6} & 0.55 \\
{[}1500, 2000) & \multicolumn{1}{c}{0.65} & 0.6 & \multicolumn{1}{c}{0.5} & 0.5 & \multicolumn{1}{c}{0.7} & 0.5 \\
{[}2000, $+\infty$) & \multicolumn{1}{c}{0.7} & 0.65 & \multicolumn{1}{c}{0.6} & 0.6 & \multicolumn{1}{c}{0.75} & 0.6 \\ \bottomrule
\end{tabular}
\caption{Division point of observable $\cosbt$ (defined in \autoref{eq:cosbartheta}) 
for different processes and collision energies. 
For each listed invariant mass bin, the events are further divided into two bins with $\cosbt<x$ and $\cosbt>x$ where we denote $x$ as the division point.
}
\label{tab:divide}
\end{table}

For the asymmetric processes ($\mu^-\mu^+ \to c s \nu_{\mu} \mu$,~ $\mu^-\mu^+ \to \tau \nu_\tau \nu_{\mu} \mu$), we consider only the measurement of the total cross section and combine its $\chi^2$ with that of the corresponding symmetric process. This also brings nontrivial improvement on the top of the symmetric process, as shown in the next section.  
Note that, similar to the symmetric processes, here the events with high invariant masses are also potentially subject to EFT validity issues.  For $\mu^-\mu^+ \to \tau \nu_\tau \nu_{\mu} \mu$ in particular, it is nontrivial to remove these events since $M_{\tau \nu_\tau}$ cannot be directly measured.  Nevertheless, since we only use the total rates of the asymmetric processes, we do not expect the events with high invariant masses to have a significant impact in our study.  A more careful treatment on the differential distributions of the asymmetric processes is left for future studies.

\section{Results}
\label{sec:results}

Our fit results are presented in this section.  In \autoref{subsec:invbin} and \ref{subsec:cosbin}, we consider the impacts of binning in the invariant mass and $\cosbt$, using $\mu^{-}\mu^{+}\rightarrow b\bar{b}\nu_{\mu}\bar{\nu}_{\mu}$ as an example.  In \autoref{subsec:unpair}, we compare the results from the $\mu^{-}\mu^{+}\rightarrow \tau^{-}\tau^{+}\nu_{\mu}\bar{\nu}_{\mu}$ process alone with the ones that also include the $ \tau \nu_\tau \nu_{\mu} \mu$ process, where the latter brings a sizable improvement.  Finally, a summary of the reaches on all Wilson coefficients is provided in \autoref{subsec:wc}.  
Note that, for the results in \autoref{subsec:invbin}, \ref{subsec:cosbin} and \ref{subsec:unpair}, we have omitted the flavor indices of the Wilson coefficients, which are $33$ (diagonal, third generation).  The flavor indices are restored in \autoref{subsec:wc}.  
Additional results are provided in \ref{app:delta_rho}.

\subsection{Impacts of the invariant mass bins}
\label{subsec:invbin}

\begin{figure}[t]
\centering
\subfigure{
    \includegraphics[width=0.30\textwidth]{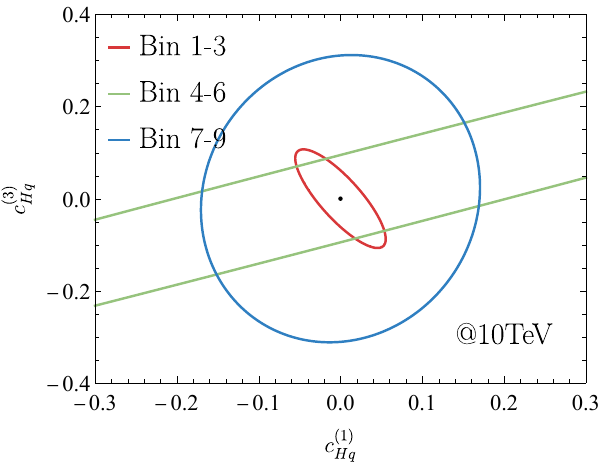}
}
\subfigure{
    \includegraphics[width=0.30\textwidth]{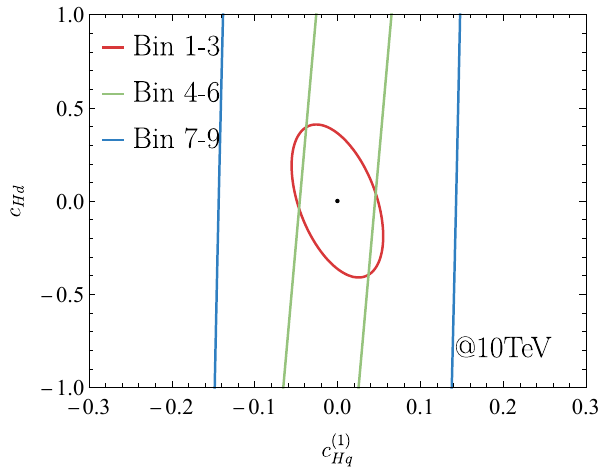}
}
\subfigure{
    \includegraphics[width=0.30\textwidth]{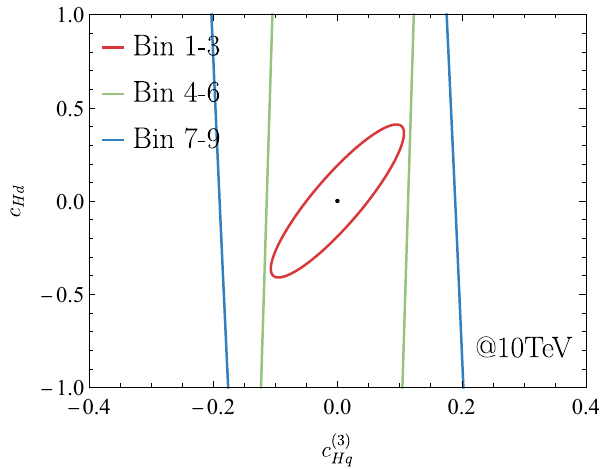}
}
\newline
\subfigure{
    \includegraphics[width=0.30\textwidth]{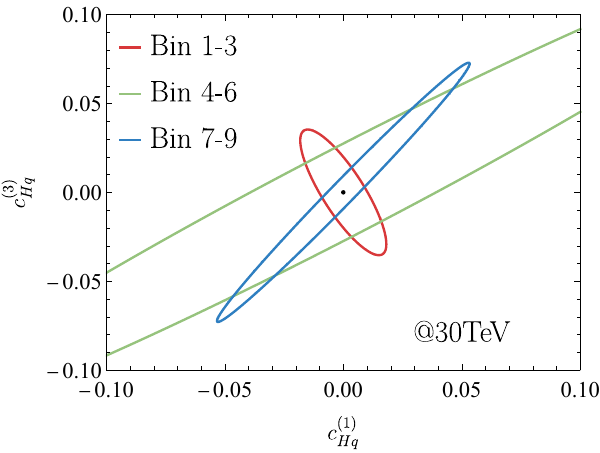}
}
\subfigure{
    \includegraphics[width=0.30\textwidth]{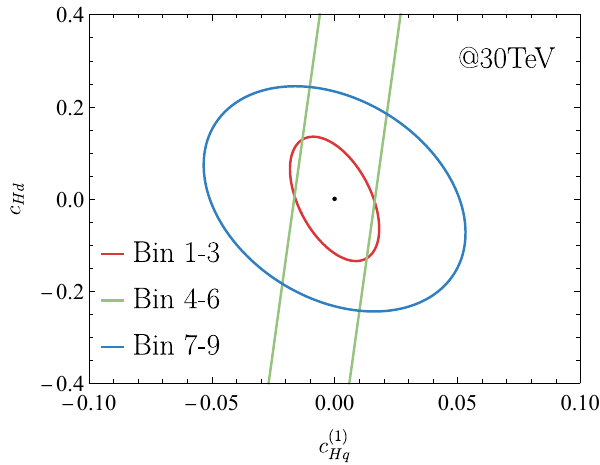}
}
\subfigure{
    \includegraphics[width=0.30\textwidth]{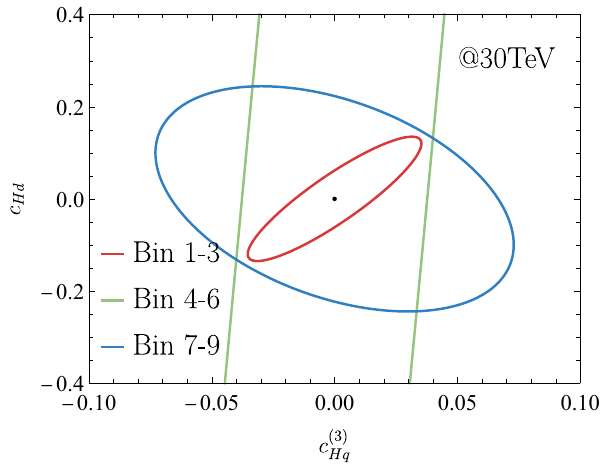}
}
\caption{The $\Delta\chi^2=1$ contours from the 3-parameter $(c_{Hq}^{(1)},c_{Hq}^{(3)},c_{Hd})$ fit to $\mu^{-}\mu^{+}\rightarrow b\bar{b}\nu_{\mu}\bar{\nu}_{\mu}$ at 10\,TeV (top row) and 30\,TeV (bottom row), obtained from three separate groups of invariant mass bins as listed in \autoref{tab:invbin}, which are bin 1-3 (red contours), bin 4-6 (green contours) and bin 7-9 (blue contours).   
For each row, the results are projected onto three 2D planes (each with the other parameter marginalized). 
We set $\Lambda=1\,$TeV for convenience. 
}
\label{fig:inv_bb_bingroup}
\end{figure}

\begin{figure}[t]
\centering
\subfigure{
    \includegraphics[width=0.30\textwidth]{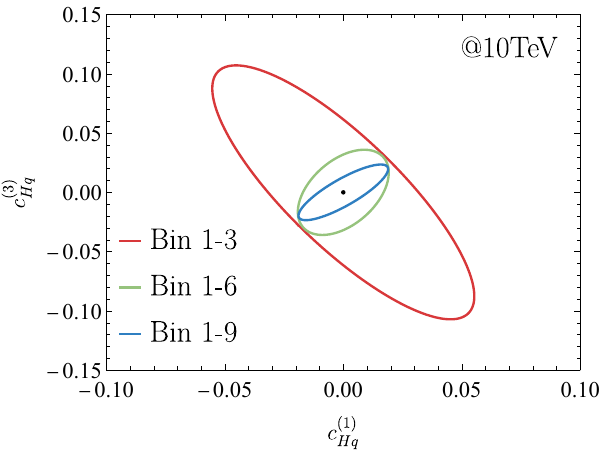}
}
\subfigure{
    \includegraphics[width=0.30\textwidth]{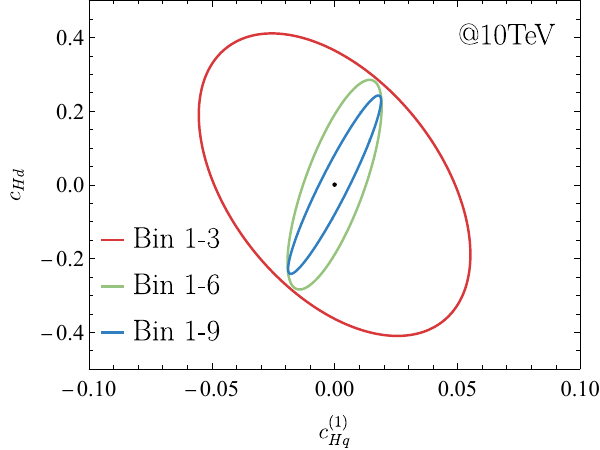}
}
\subfigure{
    \includegraphics[width=0.30\textwidth]{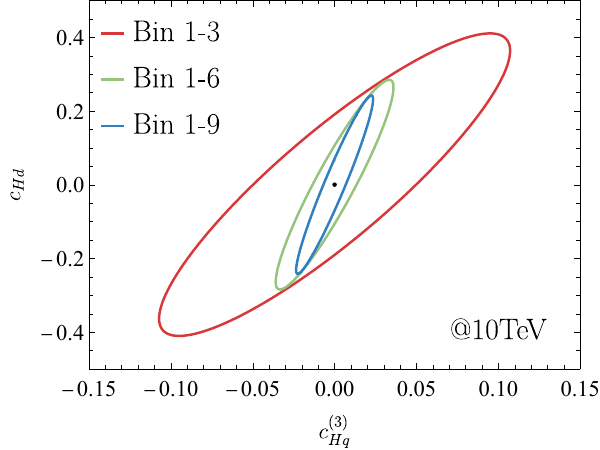}
}
\newline
\subfigure{
    \includegraphics[width=0.30\textwidth]{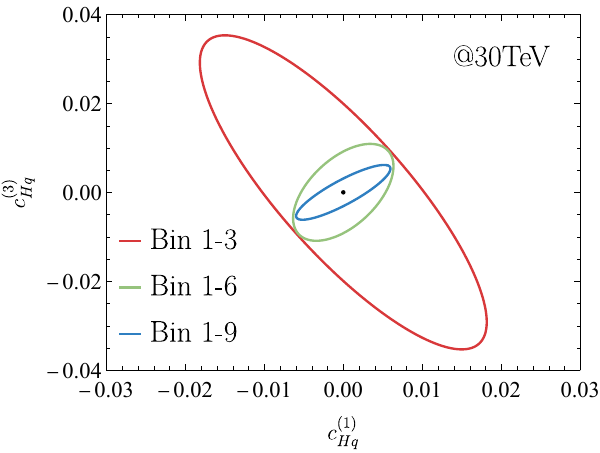}
}
\subfigure{
    \includegraphics[width=0.30\textwidth]{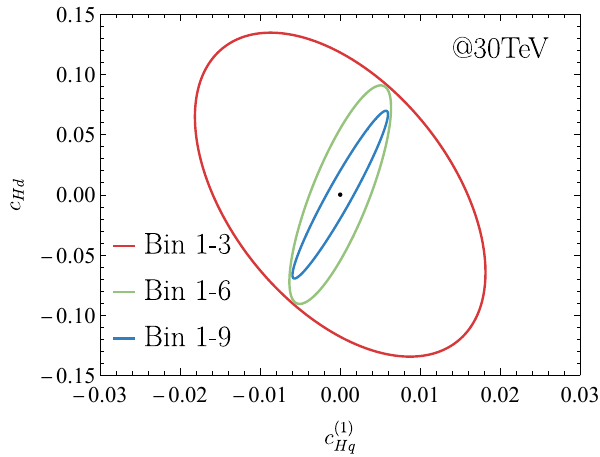}
}
\subfigure{
    \includegraphics[width=0.30\textwidth]{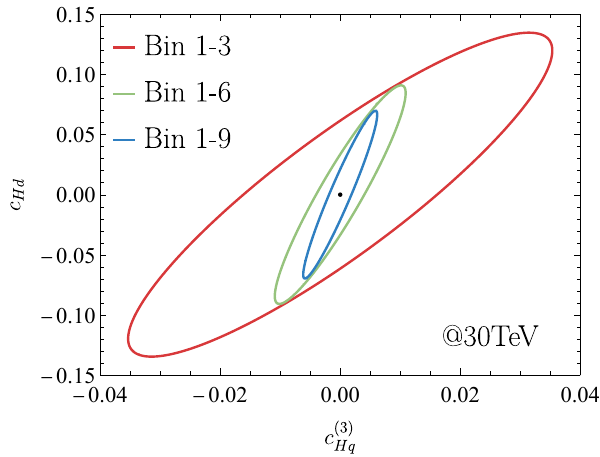}
}
\caption{The $\Delta\chi^2=1$ contours from the 3-parameter $(c_{Hq}^{(1)},c_{Hq}^{(3)},c_{Hd})$ fit 
to $\mu^{-}\mu^{+}\rightarrow b\bar{b}\nu_{\mu}\bar{\nu}_{\mu}$ at 10\,TeV (top row) and 30\,TeV (bottom row), with the invariant mass bins in \autoref{tab:invbin}.  For each row, the results are projected onto three 2D planes (each with the other parameter marginalized). The contours with the first 3, 6, and all 9 bins are shown. We set $\Lambda=1\,$TeV for convenience. 
}
\label{fig:inv_bb}
\end{figure}

To illustrate the impacts of different invariant mass bins, 
we show in \autoref{fig:inv_bb_bingroup} the $\Delta\chi^2=1$ contours
\footnote{Note that $\Delta\chi^2=\chi^2 - \chi^2_{\rm min}$, where $\chi^2_{\rm min}=\chi^2(c_i=0)=0$ by our assumption that all measurements are SM like.} 
of 3-parameter $(c_{Hq}^{(1)},c_{Hq}^{(3)},c_{Hd})$ fit of the process $\mu^{-}\mu^{+}\rightarrow b\bar{b}\nu_{\mu}\bar{\nu}_{\mu}$ for both the 10\,TeV run (top row) and the 30\,TeV run (bottom row).  In each case, we divide the invariant mass bins into three groups (bin 1-3, 4-6, and 7-9), and the $\Delta\chi^2=1$ contours are shown separately for each group of 3 bins, which are projected on three different 2D planes of the 3 parameters.   
The $\cosbt$ variable is not considered here.  
Note that, without binning (which is essentially the same as including only the first bin since the cross section is dominated by the $Z$ resonance), the cross section measurement only provides one constraint on the 3 parameters.  
A combination of at least three different bins are required to simultaneously constrain all three parameters.  Among the 3 groups of bins, clearly bin 1-3 provides the best constraints due to the large rates around the $Z$ pole.  The results from higher invariant mass bins alone are general worse, and they also suffer from large (approximate) flat directions. Nevertheless, the high invariant mass bins carry important 
complementary information to those of the $Z$-resonance events.  
This is made more clear in \autoref{fig:inv_bb}, where we present the same set of results, but for the combinations of bin 1-3, bin 1-6 and bin 1-9 instead, which clearly demonstrates the improvements brought by the higher invariant mass bins on the top of the ones around the $Z$ pole.  As mentioned earlier, we will discard the signal events with invariant mass larger than 1\,TeV (bin 7-9) in order to reduce background and improve EFT validity.  While this would reduce the overall reach, one could also see in \autoref{fig:inv_bb} that the impact of this cut is under control given that reaches with bin 1-6 (green contours) are not much worse than the ones with all the bins (blue contours).     

The results for $\mu^-\mu^+ \to c \bar{c}\nu_{\mu}\bar{\nu}_{\mu}$ and $\mu^-\mu^+ \to \tau^- \tau^+\nu_{\mu}\bar{\nu}_{\mu}$ are shown in \autoref{fig:inv_cc} and \autoref{fig:inv_tata} in \ref{app:delta_rho}, respectively, which exhibit similar features.

\subsection{Impacts of the $\cosbt$ bins}
\label{subsec:cosbin}

\begin{figure}[htb]
\centering
\subfigure{
    \includegraphics[width=0.30\textwidth]{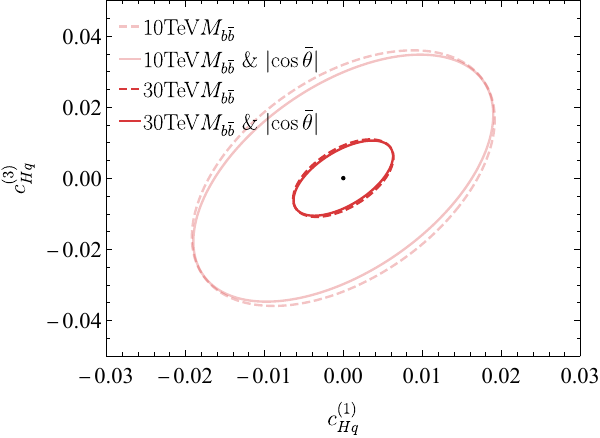}
    \label{subfig:chq1chq3}
}
\subfigure{
    \includegraphics[width=0.30\textwidth]{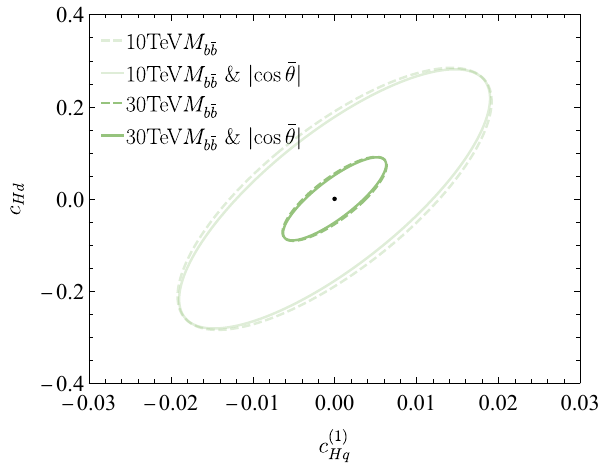}
    \label{subfig:chq1chd}
}
\subfigure{
    \includegraphics[width=0.30\textwidth]{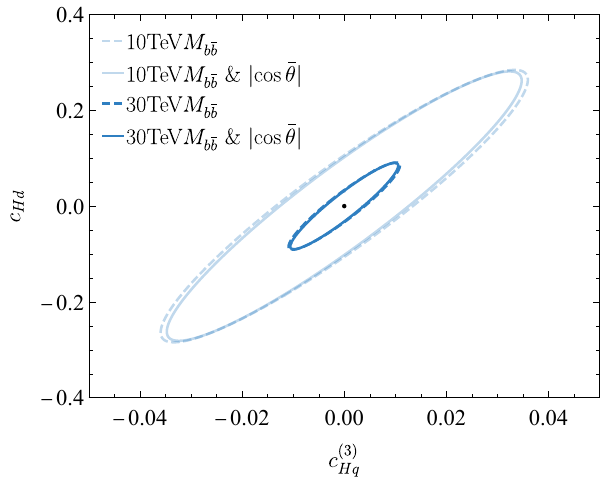}
    \label{subfig:chq3chd}
}
\caption{Comparison of the results for  $\mu^{-}\mu^{+} \to b\bar{b}\nu_{\mu}\bar{\nu}_{\mu}$ with only the invariant mass bins (dashed contours, binning as in \autoref{tab:invbin}) and also with the $\cosbt$ bins (solid contours, binning as in \autoref{tab:divide}). An invariant mass cut of $M_{b\bar{b}}<1$\,TeV is imposed. The contours correspond to $\Delta\chi^2=1$ from the 3-parameter $(c_{Hq}^{(1)},c_{Hq}^{(3)},c_{Hd})$ fit.  We set $\Lambda=1\,$TeV for convenience.
}
\label{fig:cosbin_bb}
\end{figure}

In addition to the invariant mass bins,
we consider the impacts of further splitting the high invariant mass bins ($M_{b\bar{b}}>600$\,GeV) in  $\cosbt$ as in \autoref{tab:divide}.  In \autoref{fig:cosbin_bb}, a comparison is made between the results that only include invariant mass bins ($M_{b\bar{b}}$, dashed contours) and the results that further split the bins in $\cosbt$ ($M_{b\bar{b}} ~\&~ \cosbt$, solid contours). Although the improvements from the binning in $\cosbt$ are relatively small in all cases, they are still visible, especially for the 10\,TeV case. 
In \ref{app:delta_rho}, similar results can be found in \autoref{fig:cosbin_cc} for $\mu^{-}\mu^{+}\rightarrow c\bar{c}\nu_{\mu}\bar{\nu}_{\mu}$ and \autoref{fig:cosbin_tata} for $\mu^{-}\mu^{+}\rightarrow \tau^- \tau^+ \nu_{\mu}\bar{\nu}_{\mu}$.  

\subsection{Impacts of the asymmetric processes}
\label{subsec:unpair}

\begin{figure}[htb]
\centering
\subfigure{
    \includegraphics[width=0.30\textwidth]{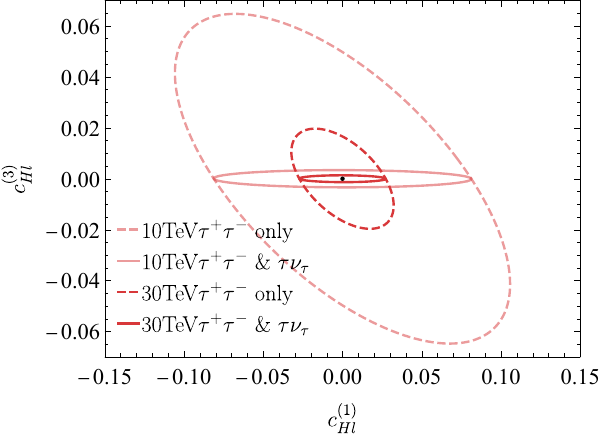}
    \label{subfig:chl1chl3}
}
\subfigure{
    \includegraphics[width=0.30\textwidth]{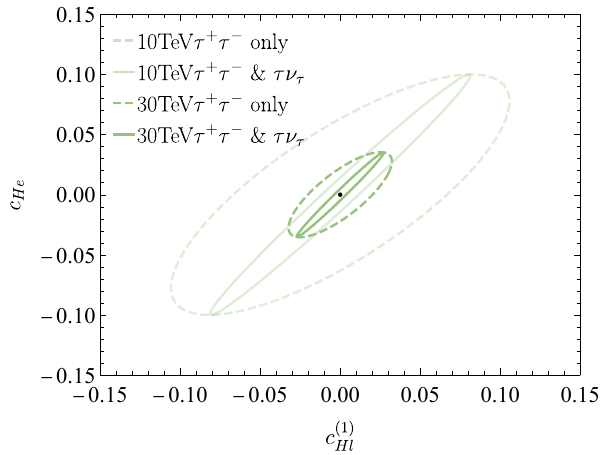}
    \label{subfig:chl1che}
}
\subfigure{
    \includegraphics[width=0.30\textwidth]{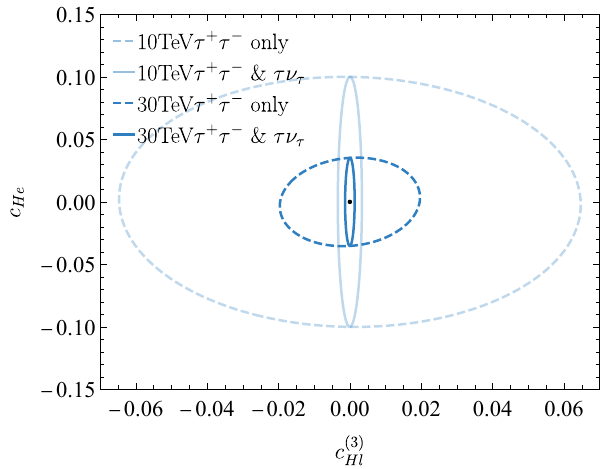}
    \label{subfig:chl3che}
}
\caption{Comparison of the results for the three parameters $c_{Hl}^{(1)}, c_{Hl}^{(3)}, c_{He}$ (all with flavor indices 33) from the measurements of $\mu^-\mu^+ \to \tau^- \tau^+\nu_{\mu}\bar{\nu}_{\mu}$ only (labeled as ``$\tau^+ \tau^-$ only'', dashed contours, with $M_{\tau^+\tau^-}<1\,$TeV) and the combination of the $\mu^-\mu^+ \to \tau^- \tau^+\nu_{\mu}\bar{\nu}_{\mu}$ and $\mu^-\mu^+ \to \tau \nu_\tau \nu_{\mu} \mu$ (labeled as ``$\tau^+ \tau^- \& \tau\nu_\tau$'', solid contours).  Contours correspond to $\Delta\chi^2=1$.  We set $\Lambda=1\,$TeV for convenience.
}
\label{fig:unpair_tata}
\end{figure}

On the top of the full binned analysis of the symmetric process, the inclusion of the asymmetric process may bring additional improvement. Note that in our study we do not consider the asymmetric process of $\mu^-\mu^+ \to b \bar{b}\nu_{\mu}\bar{\nu}_{\mu}$, which is $\mu^-\mu^+ \to tb\mu\nu_\mu$, due to the complicated nature of the top decay.  Instead, a comparison is made for $\mu^-\mu^+ \to \tau^- \tau^+\nu_{\mu}\bar{\nu}_{\mu}$ and $\mu^-\mu^+ \to \tau \nu_\tau \nu_{\mu} \mu$, with the results shown in \autoref{fig:unpair_tata}.  The combination of the two processes ($\tau^+\tau^- ~\&~ \tau \nu_\tau$, solid contour) brings a sizable improvement to the top of the symmetric process alone ($\tau^+\tau^-$ only, dashed contour), especially for the 10\,TeV run.  The improvement is most significant for $c_{H l}^{(3)}$, which modifies the $W$-fermion couplings.   
As mentioned in \autoref{subsec:method}, we have imposed a rapidity cut of $|\eta_\mu|<6$ on muons, assuming forward muon taggers~\cite{Ruhdorfer:2023uea, Ruhdorfer:2024dgz, Li:2024joa} will be implemented. Without forward muon taggers, the asymmetric processes could suffer from a much smaller signal selection efficiency.  Our analysis thus provides an important case that motivates the implementation of forward muon taggers.

Similar results for the combination of the processes $\mu^-\mu^+ \to c \bar{c}\nu_{\mu}\bar{\nu}_{\mu}$ and $\mu^-\mu^+ \to c s \nu_{\mu} \mu$ can be found in \autoref{fig:unpair_cc} in \ref{app:delta_rho}, 
which also exhibits significant improvement for $c_{H q}^{(3)}$.    

\subsection{Constraints on the Wilson coefficients}
\label{subsec:wc}

\begin{figure}[htb]
\centering
\includegraphics[width=0.8\textwidth]{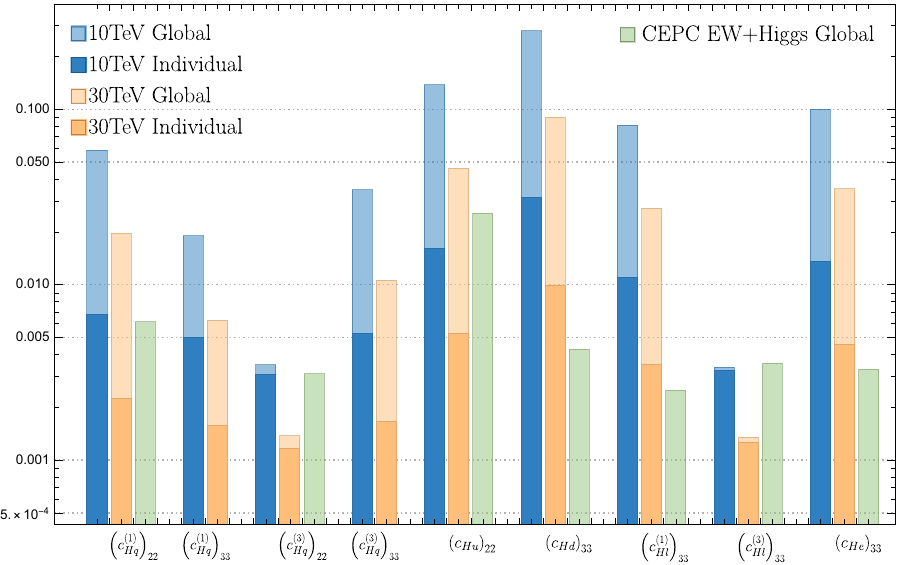}
\caption{The 1 $\sigma$ constraints on Wilson coefficients in our analysis. 
The light-shaded bars correspond to global-fit results, while the solid bars are individual fit results.  The blue (orange) bars correspond to the results at the 10\,TeV (30\,TeV) run.  The CEPC results from a full EW+Higgs global fit is shown in green bars, which are converted from the results in Ref.~\cite{deBlas:2022ofj}. 
We set $\Lambda=1\,$TeV for convenience.
}
\label{fig:delta}
\end{figure}

A summary of the reaches on all the Wilson coefficients considered in our analysis is presented in \autoref{fig:delta}.  The measurements of all the processes in \autoref{tab:eff} are considered, while for the symmetric processes the binning is performed with both the invariant mass of the fermion pair (as in \autoref{tab:invbin}) and the $\cosbt$ variable (\autoref{tab:divide}). 
Note again that we have imposed an invariant mass cut of $M_{f\bar{f}}<1$\,TeV in order to reduce background and improve EFT validity. 
The previously omitted flavor indices are also explicitly shown in \autoref{fig:delta}.  Both the global-fit results (in our cases which involve 3 operators for any given process) and the individual ones (by switching on one operator at a time) are shown.  

For comparison, we also show in \autoref{fig:delta} the reach of CEPC from a Higgs+EW global fit, converted from the results in Ref.~\cite{deBlas:2022ofj} to the basis in Ref.~\cite{Grzadkowski:2010es}. It should be emphasized that the results in Ref.~\cite{deBlas:2022ofj} are obtained under a different framework with more parameters and measurements included (with all the other parameters marginalized here), so a direct comparison under the same conditions is not possible. Furthermore, we have excluded the CEPC $t\bar{t}$ measurements (hence the bounds on $(c^{(1)}_{Hq})_{33}$ and $(c^{(3)}_{Hq})_{33}$ are missing) since the top measurements at muon collider are not included in our analysis either. Nevertheless, it can be seen that the overall reach of the muon collider on these Wilson coefficients are generally at the same order as the ones at CEPC (or FCC-ee).  

The numerical values of the one-sigma bounds of the Wilson coefficients and their correlation matrices are listed in \autoref{tab:dr_bb}, \ref{tab:dr_cc} \& \ref{tab:dr_tata}.

\begin{table}[htb]
\flushleft
\begin{minipage}[b]{0.45\textwidth}
\centering
\scalebox{0.85}{
\begin{tabular}{@{}cccccc@{}}
\multicolumn{6}{c}{\SI{10}{TeV}} \\
\toprule
                        & \multicolumn{2}{m{3cm}<{\centering}}{68\%CL $1\sigma$ bound}         & \multicolumn{3}{c}{Correlation}                        \\
                        & \multicolumn{2}{m{3cm}<{\centering}}{($\times 10^{-2}$)}      & \multicolumn{3}{c}{matrix}                             \\
                        & \multicolumn{1}{m{2cm}<{\centering}}{Individual} & Global & \multicolumn{1}{c}{\hspace{3pt}$\left(c_{Hq}^{(1)}\right)_{33}$} & \multicolumn{1}{c}{$\left(c_{Hq}^{(3)}\right)_{33}$} & $\left(c_{Hd}\right)_{33}$ \\ \midrule
$\left(c_{Hq}^{(1)}\right)_{33}$ & $\pm$0.500                          & $\pm$1.90     & 1                       &                         &    \\
$\left(c_{Hq}^{(3)}\right)_{33}$                      & $\pm$0.526                          & $\pm$3.47     & 0.871                       & 1                       &    \\
$\left(c_{Hd}\right)_{33}$                     & $\pm$3.14                           & $\pm$28.1      & 0.954                       & 0.963                       & 1  \\ 
\bottomrule
\end{tabular}
}
\end{minipage}
\hspace{2em}
\begin{minipage}[b]{0.45\textwidth}
\centering
\scalebox{0.85}{
\begin{tabular}{@{}cccccc@{}}
\multicolumn{6}{c}{\SI{30}{TeV}} \\
\toprule
                        & \multicolumn{2}{m{3cm}<{\centering}}{68\%CL $1\sigma$ bound}         & \multicolumn{3}{c}{Correlation}                        \\
                        & \multicolumn{2}{m{3cm}<{\centering}}{($\times 10^{-2}$)}      & \multicolumn{3}{c}{matrix}                             \\
                        & \multicolumn{1}{m{2cm}<{\centering}}{Individual} & Global & \multicolumn{1}{c}{\hspace{3pt}$\left(c_{Hq}^{(1)}\right)_{33}$} & \multicolumn{1}{c}{$\left(c_{Hq}^{(3)}\right)_{33}$} & $\left(c_{Hd}\right)_{33}$ \\ \midrule
$\left(c_{Hq}^{(1)}\right)_{33}$ & $\pm$0.158                          & $\pm$0.629     & 1                        &                         &    \\
$\left(c_{Hq}^{(3)}\right)_{33}$                      & $\pm$0.165                          & $\pm$1.06     & 0.895                       & 1                       &    \\
$\left(c_{Hd}\right)_{33}$                      & $\pm$0.992                           & $\pm$9.01      & 0.966                       & 0.961                       & 1  \\ 
\bottomrule
\end{tabular}
}
\end{minipage}
\caption{One-sigma individual and global bounds of $\left(c_{Hq}^{(1)}\right)_{33}$, $\left(c_{Hq}^{(3)}\right)_{33}$ and $\left(c_{Hd}\right)_{33}$ and their correlations (corresponding to the global bounds) from the measurement of $\mu^{-}\mu^{+}\rightarrow b\bar{b}\nu_{\mu}\bar{\nu}_{\mu}$ at the \SI{10}{TeV} run (left panel) and \SI{30}{TeV} run (right panel). An invariant mass cut of $M_{f\bar{f}}<1$\,TeV is imposed, after which all the invariant mass and $\cosbt$ bins are included.  
We set $\Lambda=1\,$TeV for convenience. 
}
\label{tab:dr_bb}
\end{table}

\begin{table}[htb]
\flushleft
\begin{minipage}[b]{0.45\textwidth}
\centering
\scalebox{0.85}{
\begin{tabular}{@{}cccccc@{}}
\multicolumn{6}{c}{\SI{10}{TeV}} \\
\toprule
                        & \multicolumn{2}{m{3cm}<{\centering}}{68\%CL $1\sigma$ bound}         & \multicolumn{3}{c}{Correlation}                        \\
                        & \multicolumn{2}{m{3cm}<{\centering}}{($\times 10^{-2}$)}      & \multicolumn{3}{c}{matrix}                             \\
                        & \multicolumn{1}{m{2cm}<{\centering}}{Individual} & Global & \multicolumn{1}{c}{\hspace{3pt}$\left(c_{Hq}^{(1)}\right)_{22}$} & \multicolumn{1}{c}{$\left(c_{Hq}^{(3)}\right)_{22}$} & $\left(c_{Hu}\right)_{22}$ \\ \midrule
$\left(c_{Hq}^{(1)}\right)_{22}$ & $\pm$0.679                          & $\pm$5.86     & 1                       &                         &    \\
$\left(c_{Hq}^{(3)}\right)_{22}$                      & $\pm$0.310                          & $\pm$0.349     & -0.142                       & 1                       &    \\
$\left(c_{Hu}\right)_{22}$                     & $\pm$1.61                           & $\pm$13.8      & 0.907                       & -0.323                       & 1  \\ 
\bottomrule
\end{tabular}
}
\end{minipage}
\hspace{2em}
\begin{minipage}[b]{0.45\textwidth}
\centering
\scalebox{0.85}{
\begin{tabular}{@{}cccccc@{}}
\multicolumn{6}{c}{\SI{30}{TeV}} \\
\toprule
                        & \multicolumn{2}{m{3cm}<{\centering}}{68\%CL $1\sigma$ bound}         & \multicolumn{3}{c}{Correlation}                        \\
                        & \multicolumn{2}{m{3cm}<{\centering}}{($\times 10^{-2}$)}      & \multicolumn{3}{c}{matrix}                             \\
                        & \multicolumn{1}{m{2cm}<{\centering}}{Individual} & Global & \multicolumn{1}{c}{\hspace{3pt}$\left(c_{Hq}^{(1)}\right)_{22}$} & \multicolumn{1}{c}{$\left(c_{Hq}^{(3)}\right)_{22}$} & $\left(c_{Hu}\right)_{22}$ \\ \midrule
$\left(c_{Hq}^{(1)}\right)_{22}$ & $\pm$0.224                          & $\pm$1.97     & 1                        &                         &    \\
$\left(c_{Hq}^{(3)}\right)_{22}$                      & $\pm$0.117                          & $\pm$0.138     & -0.765                       & 1                       &    \\
$\left(c_{Hu}\right)_{22}$                      & $\pm$0.526                           & $\pm$4.61      & 0.701                       & -0.747                       & 1  \\ 
\bottomrule
\end{tabular}
}
\end{minipage}
\caption{Same as \autoref{tab:dr_bb} but for $\left(c_{Hq}^{(1)}\right)_{22}$, $\left(c_{Hq}^{(3)}\right)_{22}$ and $\left(c_{Hu}\right)_{22}$ from the measurements of $\mu^-\mu^+ \to c \bar{c}\nu_{\mu}\bar{\nu}_{\mu}$ 
 and $\mu^-\mu^+ \to c s \nu_{\mu} \mu$.  
}
\label{tab:dr_cc}
\end{table}

\begin{table}[htb]
\flushleft
\begin{minipage}[b]{0.45\textwidth}
\centering
\scalebox{0.85}{
\begin{tabular}{@{}cccccc@{}}
\multicolumn{6}{c}{\SI{10}{TeV}} \\
\toprule
                        & \multicolumn{2}{m{3cm}<{\centering}}{68\%CL $1\sigma$ bound}         & \multicolumn{3}{c}{Correlation}                        \\
                        & \multicolumn{2}{m{3cm}<{\centering}}{($\times 10^{-2}$)}      & \multicolumn{3}{c}{matrix}                             \\
                        & \multicolumn{1}{m{2cm}<{\centering}}{Individual} & Global & \multicolumn{1}{c}{\hspace{3pt}$\left(c_{Hl}^{(1)}\right)_{33}$} & \multicolumn{1}{c}{$\left(c_{Hl}^{(3)}\right)_{33}$} & $\left(c_{He}\right)_{33}$ \\ \midrule
$\left(c_{Hl}^{(1)}\right)_{33}$ & $\pm$1.10                          & $\pm$8.12     & 1                       &                         &    \\
$\left(c_{Hl}^{(3)}\right)_{33}$                      & $\pm$0.323                          & $\pm$0.338     & -0.00695                       & 1                       &    \\
$\left(c_{He}\right)_{33}$                     & $\pm$1.35                           & $\pm$10.0      & 0.986                       & 0.0431                       & 1  \\ 
\bottomrule
\end{tabular}
}
\end{minipage}
\hspace{2em}
\begin{minipage}[b]{0.45\textwidth}
\centering
\scalebox{0.85}{
\begin{tabular}{@{}cccccc@{}}
\multicolumn{6}{c}{\SI{30}{TeV}} \\
\toprule
                        & \multicolumn{2}{m{3cm}<{\centering}}{68\%CL $1\sigma$ bound}         & \multicolumn{3}{c}{Correlation}                        \\
                        & \multicolumn{2}{m{3cm}<{\centering}}{($\times 10^{-2}$)}      & \multicolumn{3}{c}{matrix}                             \\
                        & \multicolumn{1}{m{2cm}<{\centering}}{Individual} & Global & \multicolumn{1}{c}{\hspace{3pt}$\left(c_{Hl}^{(1)}\right)_{33}$} & \multicolumn{1}{c}{$\left(c_{Hl}^{(3)}\right)_{33}$} & $\left(c_{He}\right)_{33}$ \\ \midrule
$\left(c_{Hl}^{(1)}\right)_{33}$ & $\pm$0.350                          & $\pm$2.72     & 1                        &                         &    \\
$\left(c_{Hl}^{(3)}\right)_{33}$                      & $\pm$0.125                          & $\pm$0.134     & 0.0388                       & 1                       &    \\
$\left(c_{He}\right)_{33}$                      & $\pm$0.453                           & $\pm$3.53      & 0.946                       & 0.156                       & 1  \\ 
\bottomrule
\end{tabular}
}
\end{minipage}
\caption{Same as \autoref{tab:dr_bb} but for $\left(c_{Hl}^{(1)}\right)_{33}$, $\left(c_{Hl}^{(3)}\right)_{33}$ and $\left(c_{He}\right)_{33}$ from the measurements of $\mu^-\mu^+ \to \tau^- \tau^+\nu_{\mu}\bar{\nu}_{\mu}$ 
and $\mu^-\mu^+ \to \tau \nu_\tau \nu_{\mu} \mu$. 
}
\label{tab:dr_tata}
\end{table}

\section{Conclusion}
\label{sec:con}

In this paper, we considered the measurements of several vector-boson-fusion-to-two-fermions (VBF$\to 2f$, with an additional $\bar{\nu}_\mu \nu_\mu$ or $\nu_\mu \mu$ pair) processes at a future high energy muon collider, focusing on final states involving $b,c,\tau$.  
A phenomenological study is performed to estimate their potential in probing the corresponding dimension-6 operators that directly modifies the couplings of the fermions ($b,c,\tau$) to the W and Z bosons.  With realistic tagging efficiencies applied, we considered only signal statistical uncertainties and extracted the precision reaches on the Wilson coefficients with a chi-squared analysis on the binned signal distributions. The information in the invariant mass of the two-fermion pair turned out to be crucial for discriminating the effects of different operator and simultaneously constraining their coefficients in a global fit.  The symmetric ($WW$ fusion) and asymmetric ($WZ/W\gamma$ fusion) processes also contain complementary information. 
Combining all measurements, the precision on the relevant Wilson coefficients (setting $\Lambda=1\,$TeV) from a simultaneous fit with all the relevant operators in consideration ({\it i.e.} those listed in \autoref{eq:Op1} and \autoref{eq:Op2}) at a future muon collider can reach up to the $10^{-2}$ level for the 10\,TeV run and up to $\sim 5\times10^{-3}$ for the 30\,TeV run.  For the latter, the results can be competitive with the ones from a future $e^+e^-$ collider with a dedicated Z-pole run, such as the CEPC.  

Our analysis demonstrates the great potential of a future high energy muon collider in precision EW measurements, which calls for further studies in this direction.  
Ultimately, a complete analysis that includes the all the relevant EW measurements and operators are needed to fully determine the potential of a muon collider in probing EW precision physics, and our study serves as one of the many early steps towards this goal.  There are several directions for future studies.  
Most importantly, realistic analyses that include detector simulation and careful treatments of background effects and signal selection efficiencies are needed, once a more concrete detector design is available.  Given the overall good statistical precision reaches, it is also important to study the effects of systematics and theory uncertainties.  These effects could be different from the ones of EW measurements at future $e^+e^-$ colliders due to the different processes and collider environments, and may require dedicated studies.  The usefulness of differential distributions illustrated in our study is also a call for a more sophisticated analysis of the distribution, which perhaps makes use of optimal observables~\cite{Diehl:1993br} and/or machine learning technics (see {\it e.g.} Ref.~\cite{Chai:2024zyl}).  The study of the muon collider's EW physics potential is also particularly relevant for the particle physics roadmap.
If a muon collider is eventually built while a $e^+e^-$ collider is not, would we miss any important physics without a Z-pole program?  If both colliders are built, are there any important complementarity that the muon collider could offer in the EW precision physics? 
These important questions needs to be addressed when we make plans for the future colliders.

\section*{Acknowledgments }

We thank Ilaria Brivio for the prompt and helpful reply to our questions regarding the SMEFTsim package. We also thank Wantong Jiang for collaborations in the early stages of this work.  This work is supported by the National Natural Science Foundation of China (NSFC) under grant No.\,12035008, No.\,12375091 and No.\,12347171, and the Innovation Program for Quantum Science and Technology under grant No.\,2024ZD0300101. 
X.Z. Tan also acknowledges the support from Helmholtz – OCPC (Office of China Postdoctoral Council) Postdoctoral Fellowship Program.


\appendix
\renewcommand{\appendixname}{} 
\renewcommand{\thesection}{Appendix \Alph{section}} 
\section{Numerical expressions of cross section}
\label{app:xs}

In \autoref{tab:xs_bb10} to \ref{tab:xs_tata30}, 
contributions of relative Wilson coefficients to the cross section are formulated under different invariant mass bins and angle divisions of $\cosbt$, where only linear contributions are considered as mentioned in \autoref{subsec:proc}. The total cross sections corresponding to asymmetric processes without binning of invariant mass and angle divisions are also listed, when applicable.  In all cases, $\Lambda$ is set to 1\,TeV.

\begin{table}[h]
\centering
\renewcommand{\arraystretch}{1.5}
\resizebox{\textwidth}{!}{
\begin{tabular}{m{4cm}<{\centering}|m{2cm}<{\centering}|m{2cm}<{\centering}|m{10cm}<{\centering}}
\toprule
Invariant mass [GeV]\vspace{4pt} & Polar angle $\cosbt$ & SM cross section [pb] & Normalized\vspace{4pt} SMEFT cross section \\
\midrule
$[0, +\infty)$ & -- & 0.302 & $1 + 0.144 \left(c_{Hq}^{(1)}\right)_{33} + 0.131 \left(c_{Hq}^{(3)}\right)_{33} - 0.0226 \left(c_{Hd}\right)_{33}$ \\
\midrule
\midrule
$[0, 100)$ & -- & 0.275 & $1 + 0.138 \left(c_{Hq}^{(1)}\right)_{33} + 0.138 \left(c_{Hq}^{(3)}\right)_{33} - 0.0238 \left(c_{Hd}\right)_{33}$ \\
\hline
$[100, 200)$ & -- & 0.0185 & $1 + 0.137 \left(c_{Hq}^{(1)}\right)_{33} + 0.127 \left(c_{Hq}^{(3)}\right)_{33} - 0.0140 \left(c_{Hd}\right)_{33}$ \\
\hline
$[200, 400)$ & -- & $0.00405$ & $1 + 0.186 \left(c_{Hq}^{(1)}\right)_{33} + 0.0611 \left(c_{Hq}^{(3)}\right)_{33} - 0.00527 \left(c_{Hd}\right)_{33}$ \\
\hline
$[400, 600)$ & -- & $0.00182$ & $1 + 0.290 \left(c_{Hq}^{(1)}\right)_{33} - 0.0442 \left(c_{Hq}^{(3)}\right)_{33} - 0.00539 \left(c_{Hd}\right)_{33}$ \\
\hline
\multirow{2}{*}{$[600, 800)$} & $\leq 0.45$ & $1.98 \times 10^{-4}$ & $1 + 0.622 \left(c_{Hq}^{(1)}\right)_{33} - 0.267 \left(c_{Hq}^{(3)}\right)_{33} - 0.00682 \left(c_{Hd}\right)_{33}$ \\
\cline{2-4}
 & $> 0.45$ & $4.35 \times 10^{-4}$ & $1 + 0.307 \left(c_{Hq}^{(1)}\right)_{33} - 0.111 \left(c_{Hq}^{(3)}\right)_{33} - 0.00631 \left(c_{Hd}\right)_{33}$ \\
\hline
\multirow{2}{*}{$[800, 1000)$} & $\leq 0.55$ & $1.42 \times 10^{-4}$ & $1 + 0.810 \left(c_{Hq}^{(1)}\right)_{33} - 0.498 \left(c_{Hq}^{(3)}\right)_{33} - 0.00861 \left(c_{Hd}\right)_{33}$ \\
\cline{2-4}
 & $> 0.55$ & $2.25 \times 10^{-4}$ & $1 + 0.346 \left(c_{Hq}^{(1)}\right)_{33} - 0.152 \left(c_{Hq}^{(3)}\right)_{33} - 0.00749 \left(c_{Hd}\right)_{33}$ \\
\hline
\multirow{2}{*}{$[1000, 1500)$} & $\leq 0.6$ & $1.67 \times 10^{-4}$ & $1 + 1.15 \left(c_{Hq}^{(1)}\right)_{33} - 0.879 \left(c_{Hq}^{(3)}\right)_{33} - 0.0116 \left(c_{Hd}\right)_{33}$ \\
\cline{2-4}
 & $> 0.6$ & $2.61 \times 10^{-4}$ & $1 + 0.451 \left(c_{Hq}^{(1)}\right)_{33} - 0.229 \left(c_{Hq}^{(3)}\right)_{33} - 0.00958 \left(c_{Hd}\right)_{33}$ \\
\hline
\multirow{2}{*}{$[1500, 2000)$} & $\leq 0.65$ & $6.17 \times 10^{-5}$ & $1 + 1.78 \left(c_{Hq}^{(1)}\right)_{33} - 1.52 \left(c_{Hq}^{(3)}\right)_{33} - 0.0171 \left(c_{Hd}\right)_{33}$ \\
\cline{2-4}
 & $> 0.65$ & $1.02 \times 10^{-4}$ & $1 + 0.584 \left(c_{Hq}^{(1)}\right)_{33} - 0.347 \left(c_{Hq}^{(3)}\right)_{33} - 0.0131 \left(c_{Hd}\right)_{33}$ \\
\hline
\multirow{2}{*}{$[2000, +\infty)$} & $\leq 0.7$ & $6.22 \times 10^{-5}$ & $1 + 2.45 \left(c_{Hq}^{(1)}\right)_{33} - 2.09 \left(c_{Hq}^{(3)}\right)_{33} - 0.0279 \left(c_{Hd}\right)_{33}$ \\
\cline{2-4}
 & $> 0.7$ & $1.17 \times 10^{-4}$ & $1 + 0.706 \left(c_{Hq}^{(1)}\right)_{33} - 0.519 \left(c_{Hq}^{(3)}\right)_{33} - 0.0181 \left(c_{Hd}\right)_{33}$ \\
\bottomrule
\end{tabular}
}
\caption{The SM cross section and normalized SMEFT cross section ($\sigma_{\rm SMEFT}/\sigma_{\rm SM}$, as a function of Wilson coefficients) of $\mu^{-}\mu^{+}\rightarrow b\bar{b}\nu_{\mu}\bar{\nu}_{\mu}$ for different invariant mass and $\cosbt$ bins at $\sqrt{s}=10\TeV$. The first row corresponds to the unbinned total cross section. Tagging efficiencies are not applied here. 
}
\label{tab:xs_bb10}
\end{table}

\begin{table}[h]
\centering
\renewcommand{\arraystretch}{1.5}
\resizebox{\textwidth}{!}{
\begin{tabular}{m{4cm}<{\centering}|m{2cm}<{\centering}|m{2.5cm}<{\centering}|m{10cm}<{\centering}}
\toprule
Invariant mass [GeV]\vspace{4pt} & Polar angle $\cosbt$ & SM cross section [pb] & Normalized\vspace{4pt} SMEFT cross section \\
\midrule
$[0, +\infty)$ & -- & 0.336 & $1 + 0.147 \left(c_{Hq}^{(1)}\right)_{33} + 0.130 \left(c_{Hq}^{(3)}\right)_{33} - 0.0226 \left(c_{Hd}\right)_{33}$ \\
\midrule
\midrule
$[0, 100)$ & -- & 0.306 & $1 + 0.138 \left(c_{Hq}^{(1)}\right)_{33} + 0.138 \left(c_{Hq}^{(3)}\right)_{33} - 0.0237 \left(c_{Hd}\right)_{33}$ \\
\hline
$[100, 200)$ & -- & 0.0207 & $1 + 0.138 \left(c_{Hq}^{(1)}\right)_{33} + 0.127 \left(c_{Hq}^{(3)}\right)_{33} - 0.0141 \left(c_{Hd}\right)_{33}$ \\
\hline
$[200, 400)$ & -- & $0.00427$ & $1 + 0.191 \left(c_{Hq}^{(1)}\right)_{33} + 0.0665 \left(c_{Hq}^{(3)}\right)_{33} - 0.00614 \left(c_{Hd}\right)_{33}$ \\
\hline
$[400, 600)$ & -- & $0.00208$ & $1 + 0.258 \left(c_{Hq}^{(1)}\right)_{33} - 0.0462 \left(c_{Hq}^{(3)}\right)_{33} - 0.00607 \left(c_{Hd}\right)_{33}$ \\
\hline
\multirow{2}{*}{$[600, 800)$} & $\leq 0.5$ & $2.98 \times 10^{-4}$ & $1 + 0.593 \left(c_{Hq}^{(1)}\right)_{33} - 0.257 \left(c_{Hq}^{(3)}\right)_{33} - 0.00792 \left(c_{Hd}\right)_{33}$ \\
\cline{2-4}
 & $> 0.5$ & $4.52 \times 10^{-4}$ & $1 + 0.297 \left(c_{Hq}^{(1)}\right)_{33} - 0.112 \left(c_{Hq}^{(3)}\right)_{33} - 0.00706 \left(c_{Hd}\right)_{33}$ \\
\hline
\multirow{2}{*}{$[800, 1000)$} & $\leq 0.5$ & $1.62 \times 10^{-4}$ & $1 + 0.870 \left(c_{Hq}^{(1)}\right)_{33} - 0.555 \left(c_{Hq}^{(3)}\right)_{33} - 0.0103 \left(c_{Hd}\right)_{33}$ \\
\cline{2-4}
 & $> 0.5$ & $2.91 \times 10^{-4}$ & $1 + 0.383 \left(c_{Hq}^{(1)}\right)_{33} - 0.175 \left(c_{Hq}^{(3)}\right)_{33} - 0.00886 \left(c_{Hd}\right)_{33}$ \\
\hline
\multirow{2}{*}{$[1000, 1500)$} & $\leq 0.5$ & $1.72 \times 10^{-4}$ & $1 + 1.45 \left(c_{Hq}^{(1)}\right)_{33} - 1.16 \left(c_{Hq}^{(3)}\right)_{33} - 0.0157 \left(c_{Hd}\right)_{33}$ \\
\cline{2-4}
 & $> 0.5$ & $3.89 \times 10^{-4}$ & $1 + 0.523 \left(c_{Hq}^{(1)}\right)_{33} - 0.294 \left(c_{Hq}^{(3)}\right)_{33} - 0.0121 \left(c_{Hd}\right)_{33}$ \\
\hline
\multirow{2}{*}{$[1500, 2000)$} & $\leq 0.6$ & $9.21 \times 10^{-5}$ & $1 + 2.14 \left(c_{Hq}^{(1)}\right)_{33} - 1.86 \left(c_{Hq}^{(3)}\right)_{33} - 0.0248 \left(c_{Hd}\right)_{33}$ \\
\cline{2-4}
 & $> 0.6$ & $1.46 \times 10^{-4}$ & $1 + 0.693 \left(c_{Hq}^{(1)}\right)_{33} - 0.460 \left(c_{Hq}^{(3)}\right)_{33} - 0.0178 \left(c_{Hd}\right)_{33}$ \\
\hline
\multirow{2}{*}{$[2000, +\infty)$} & $\leq 0.65$ & $1.30 \times 10^{-4}$ & $1 + 2.10 \left(c_{Hq}^{(1)}\right)_{33} - 1.70 \left(c_{Hq}^{(3)}\right)_{33} - 0.0749 \left(c_{Hd}\right)_{33}$ \\
\cline{2-4}
 & $> 0.$ & $2.21 \times 10^{-4}$ & $1 + 0.571 \left(c_{Hq}^{(1)}\right)_{33} - 0.420 \left(c_{Hq}^{(3)}\right)_{33} - 0.0364 \left(c_{Hd}\right)_{33}$ \\
\bottomrule
\end{tabular}
}
\caption{Same as \autoref{tab:xs_bb10} but for $\sqrt{s}=30\TeV$.
}
\label{tab:xs_bb30}
\end{table}

\begin{table}[h]
\centering
\renewcommand{\arraystretch}{1.5}
\resizebox{\textwidth}{!}{
\begin{tabular}{m{4cm}<{\centering}|m{2cm}<{\centering}|m{2cm}<{\centering}|m{10cm}<{\centering}}
\toprule
Invariant mass [GeV]\vspace{4pt} & Polar angle $\cosbt$ & SM cross section [pb] & Normalized\vspace{4pt} SMEFT cross section \\
\midrule
$[0, +\infty)$ & -- & 0.216 & $1 - 0.142 \left(c_{Hq}^{(1)}\right)_{22} + 0.148 \left(c_{Hq}^{(3)}\right)_{22} + 0.0604 \left(c_{Hu}\right)_{22}$ \\
\midrule
\midrule
$[0, 100)$ & -- & 0.199 & $1 - 0.144 \left(c_{Hq}^{(1)}\right)_{22} + 0.146 \left(c_{Hq}^{(3)}\right)_{22} + 0.0622 \left(c_{Hu}\right)_{22}$ \\
\hline
$[100, 200)$ & -- & 0.01395 & $1 - 0.140 \left(c_{Hq}^{(1)}\right)_{22} + 0.139 \left(c_{Hq}^{(3)}\right)_{22} + 0.0363 \left(c_{Hu}\right)_{22}$ \\
\hline
$[200, 400)$ & -- & $0.00252$ & $1 - 0.0823 \left(c_{Hq}^{(1)}\right)_{22} + 0.164 \left(c_{Hq}^{(3)}\right)_{22} + 0.0176 \left(c_{Hu}\right)_{22}$ \\
\hline
$[400, 600)$ & -- & $8.29 \times 10^{-4}$ & $1 - 0.0294 \left(c_{Hq}^{(1)}\right)_{22} + 0.210 \left(c_{Hq}^{(3)}\right)_{22} + 0.0247 \left(c_{Hu}\right)_{22}$ \\
\hline
\multirow{2}{*}{$[600, 800)$} & $\leq 0.45$ & $1.04 \times 10^{-4}$ & $1 + 0.0376 \left(c_{Hq}^{(1)}\right)_{22} + 0.373 \left(c_{Hq}^{(3)}\right)_{22} + 0.0488 \left(c_{Hu}\right)_{22}$ \\
\cline{2-4}
 & $> 0.45$ & $2.86 \times 10^{-4}$ & $1 + 0.0141 \left(c_{Hq}^{(1)}\right)_{22} + 0.218 \left(c_{Hq}^{(3)}\right)_{22} + 0.0292 \left(c_{Hu}\right)_{22}$ \\
\hline
\multirow{2}{*}{$[800, 1000)$} & $\leq 0.45$ & $5.38 \times 10^{-5}$ & $1 + 0.152 \left(c_{Hq}^{(1)}\right)_{22} + 0.496 \left(c_{Hq}^{(3)}\right)_{22} + 0.0684 \left(c_{Hu}\right)_{22}$ \\
\cline{2-4}
 & $> 0.45$ & $1.58 \times 10^{-4}$ & $1 + 0.0513 \left(c_{Hq}^{(1)}\right)_{22} + 0.252 \left(c_{Hq}^{(3)}\right)_{22} + 0.0368 \left(c_{Hu}\right)_{22}$ \\
\hline
\multirow{2}{*}{$[1000, 1500)$} & $\leq 0.45$ & $5.79 \times 10^{-5}$ & $1 + 0.402 \left(c_{Hq}^{(1)}\right)_{22} + 0.750 \left(c_{Hq}^{(3)}\right)_{22} + 0.104 \left(c_{Hu}\right)_{22}$ \\
\cline{2-4}
 & $> 0.45$ & $1.72 \times 10^{-4}$ & $1 + 0.129 \left(c_{Hq}^{(1)}\right)_{22} + 0.330 \left(c_{Hq}^{(3)}\right)_{22} + 0.0493 \left(c_{Hu}\right)_{22}$ \\
\hline
\multirow{2}{*}{$[1500, 2000)$} & $\leq 0.5$ & $2.51 \times 10^{-5}$ & $1 + 0.956 \left(c_{Hq}^{(1)}\right)_{22} + 1.29 \left(c_{Hq}^{(3)}\right)_{22} + 0.158 \left(c_{Hu}\right)_{22}$ \\
\cline{2-4}
 & $> 0.5$ & $5.48 \times 10^{-5}$ & $1 + 0.305 \left(c_{Hq}^{(1)}\right)_{22} + 0.495 \left(c_{Hq}^{(3)}\right)_{22} + 0.0699 \left(c_{Hu}\right)_{22}$ \\
\hline
\multirow{2}{*}{$[2000, +\infty)$} & $\leq 0.6$ & $3.63 \times 10^{-5}$ & $1 + 3.36 \left(c_{Hq}^{(1)}\right)_{22} + 3.66 \left(c_{Hq}^{(3)}\right)_{22} + 0.234 \left(c_{Hu}\right)_{22}$ \\
\cline{2-4}
 & $> 0.6$ & $4.27 \times 10^{-5}$ & $1 + 0.980 \left(c_{Hq}^{(1)}\right)_{22} + 1.14 \left(c_{Hq}^{(3)}\right)_{22} + 0.0992 \left(c_{Hu}\right)_{22}$ \\
\midrule
\midrule
$\mu^-\mu^+\rightarrow c s \nu_{\mu} \mu$ & $-$ & 0.0144 & $1 - 1.92\times10^{-4} \left(c_{Hq}^{(1)}\right)_{22} - 0.121 \left(c_{Hq}^{(3)}\right)_{22} - 7.16 \times 10^{-11} \left(c_{Hu}\right)_{22}$ \\
\bottomrule
\end{tabular}
}
\caption{
The SM cross section and normalized SMEFT cross section ($\sigma_{\rm SMEFT}/\sigma_{\rm SM}$, as a function of Wilson coefficients) of $\mu^{-}\mu^{+}\rightarrow c\bar{c}\nu_{\mu}\bar{\nu}_{\mu}$ for different invariant mass and $\cosbt$ bins at $\sqrt{s}=10\TeV$. The first (last) row corresponds to the unbinned total cross section for $\mu^{-}\mu^{+}\rightarrow c\bar{c}\nu_{\mu}\bar{\nu}_{\mu}$ ($\mu^-\mu^+ \to c s \nu_{\mu} \mu$). Tagging efficiencies are not applied here. 
}
\label{tab:xs_cc10}
\end{table}

\begin{table}[h]
\centering
\renewcommand{\arraystretch}{1.5}
\resizebox{\textwidth}{!}{
\begin{tabular}{m{4cm}<{\centering}|m{2cm}<{\centering}|m{2cm}<{\centering}|m{10cm}<{\centering}}
\toprule
Invariant mass [GeV]\vspace{4pt} & Polar angle $\cosbt$ & SM cross section [pb] & Normalized\vspace{4pt} SMEFT cross section \\
\midrule
$[0, +\infty)$ & -- & 0.225 & $1 - 0.133 \left(c_{Hq}^{(1)}\right)_{22} + 0.155 \left(c_{Hq}^{(3)}\right)_{22} + 0.0600 \left(c_{Hu}\right)_{22}$ \\
\midrule
\midrule
$[0, 100)$ & -- & 0.206 & $1 - 0.143 \left(c_{Hq}^{(1)}\right)_{22} + 0.146 \left(c_{Hq}^{(3)}\right)_{22} + 0.0622 \left(c_{Hu}\right)_{22}$ \\
\hline
$[100, 200)$ & -- & 0.0138 & $1 - 0.136 \left(c_{Hq}^{(1)}\right)_{22} + 0.137 \left(c_{Hq}^{(3)}\right)_{22} + 0.0367 \left(c_{Hu}\right)_{22}$ \\
\hline
$[200, 400)$ & -- & $0.00261$ & $1 - 0.0980 \left(c_{Hq}^{(1)}\right)_{22} + 0.184 \left(c_{Hq}^{(3)}\right)_{22} + 0.0174 \left(c_{Hu}\right)_{22}$ \\
\hline
$[400, 600)$ & -- & $0.00105$ & $1 - 0.0371 \left(c_{Hq}^{(1)}\right)_{22} + 0.204 \left(c_{Hq}^{(3)}\right)_{22} + 0.0249 \left(c_{Hu}\right)_{22}$ \\
\hline
\multirow{2}{*}{$[600, 800)$} & $\leq 0.45$ & $1.44 \times 10^{-4}$ & $1 + 0.0138 \left(c_{Hq}^{(1)}\right)_{22} + 0.357 \left(c_{Hq}^{(3)}\right)_{22} + 0.0471 \left(c_{Hu}\right)_{22}$ \\
\cline{2-4}
 & $> 0.45$ & $3.95 \times 10^{-4}$ & $1 + 0.00529 \left(c_{Hq}^{(1)}\right)_{22} + 0.210 \left(c_{Hq}^{(3)}\right)_{22} + 0.0284 \left(c_{Hu}\right)_{22}$ \\
\hline
\multirow{2}{*}{$[800, 1000)$} & $\leq 0.45$ & $7.85 \times 10^{-5}$ & $1 + 0.115 \left(c_{Hq}^{(1)}\right)_{22} + 0.489 \left(c_{Hq}^{(3)}\right)_{22} + 0.0678 \left(c_{Hu}\right)_{22}$ \\
\cline{2-4}
 & $> 0.45$ & $2.42 \times 10^{-4}$ & $1 + 0.0393 \left(c_{Hq}^{(1)}\right)_{22} + 0.236 \left(c_{Hq}^{(3)}\right)_{22} + 0.0347 \left(c_{Hu}\right)_{22}$ \\
\hline
\multirow{2}{*}{$[1000, 1500)$} & $\leq 0.5$ & $1.20 \times 10^{-4}$ & $1 + 0.320 \left(c_{Hq}^{(1)}\right)_{22} + 0.679 \left(c_{Hq}^{(3)}\right)_{22} + 0.0959 \left(c_{Hu}\right)_{22}$ \\
\cline{2-4}
 & $> 0.5$ & $2.81 \times 10^{-4}$ & $1 + 0.103 \left(c_{Hq}^{(1)}\right)_{22} + 0.298 \left(c_{Hq}^{(3)}\right)_{22} + 0.0442 \left(c_{Hu}\right)_{22}$ \\
\hline
\multirow{2}{*}{$[1500, 2000)$} & $\leq 0.5$ & $5.08 \times 10^{-5}$ & $1 + 0.919 \left(c_{Hq}^{(1)}\right)_{22} + 1.27 \left(c_{Hq}^{(3)}\right)_{22} + 0.159 \left(c_{Hu}\right)_{22}$ \\
\cline{2-4}
 & $> 0.5$ & $1.22 \times 10^{-4}$ & $1 + 0.252 \left(c_{Hq}^{(1)}\right)_{22} + 0.441 \left(c_{Hq}^{(3)}\right)_{22} + 0.0637 \left(c_{Hu}\right)_{22}$ \\
\hline
\multirow{2}{*}{$[2000, +\infty)$} & $\leq 0.6$ & $1.22 \times 10^{-4}$ & $1 + 12.1 \left(c_{Hq}^{(1)}\right)_{22} + 12.3 \left(c_{Hq}^{(3)}\right)_{22} + 0.336 \left(c_{Hu}\right)_{22}$ \\
\cline{2-4}
 & $> 0.6$ & $1.50 \times 10^{-4}$ & $1 + 2.79 \left(c_{Hq}^{(1)}\right)_{22} + 3.02 \left(c_{Hq}^{(3)}\right)_{22} + 0.113 \left(c_{Hu}\right)_{22}$ \\
\midrule
\midrule
$\mu^-\mu^+\rightarrow c s \nu_{\mu} \mu$ & $-$ & 0.520 & $1 - 3.22 \times 10^{-6} \left(c_{Hq}^{(1)}\right)_{22} + 0.126 \left(c_{Hq}^{(3)}\right)_{22} - 4.19 \times 10^{-8} \left(c_{Hu}\right)_{22}$ \\
\bottomrule
\end{tabular}
}
\caption{Same as \autoref{tab:xs_cc10} but for $\sqrt{s}=30\TeV$.
}
\label{tab:xs_cc30}
\end{table}

\begin{table}[h]
\centering
\renewcommand{\arraystretch}{1.5}
\resizebox{\textwidth}{!}{
\begin{tabular}{m{4cm}<{\centering}|m{2cm}<{\centering}|m{2cm}<{\centering}|m{10cm}<{\centering}}
\toprule
Invariant mass [GeV]\vspace{4pt} & Polar angle $\cos\bar{\theta}$ & SM cross section [pb] & Normalized\vspace{4pt} SMEFT cross section \\
\midrule
$[0, +\infty)$ & -- & 0.0740 & $1 + 0.127 \left(c_{Hl}^{(1)}\right)_{33} + 0.132 \left(c_{Hl}^{(3)}\right)_{33} - 0.102 \left(c_{He}\right)_{33}$ \\
\midrule
\midrule
$[0, 100)$ & -- & 0.0689 & $1 + 0.126 \left(c_{Hl}^{(1)}\right)_{33} + 0.127 \left(c_{Hl}^{(3)}\right)_{33} - 0.105 \left(c_{He}\right)_{33}$ \\
\hline
$[100, 200)$ & -- & 0.00452 & $1 + 0.138 \left(c_{Hl}^{(1)}\right)_{33} + 0.143 \left(c_{Hl}^{(3)}\right)_{33} - 0.0615 \left(c_{He}\right)_{33}$ \\
\hline
$[200, 400)$ & -- & $9.24 \times 10^{-4}$ & $1 + 0.0776 \left(c_{Hl}^{(1)}\right)_{33} + 0.167 \left(c_{Hl}^{(3)}\right)_{33} - 0.0241 \left(c_{He}\right)_{33}$ \\
\hline
$[400, 600)$ & -- & $3.35 \times 10^{-4}$ & $1 + 0.0338 \left(c_{Hl}^{(1)}\right)_{33} + 0.205 \left(c_{Hl}^{(3)}\right)_{33} - 0.0304 \left(c_{He}\right)_{33}$ \\
\hline
\multirow{2}{*}{$[600, 800)$} & $\leq 0.5$ & $5.34 \times 10^{-5}$ & $1 + 0.00797 \left(c_{Hl}^{(1)}\right)_{33} + 0.299 \left(c_{Hl}^{(3)}\right)_{33} - 0.0536 \left(c_{He}\right)_{33}$ \\
\cline{2-4}
& $> 0.5$ & $1.19 \times 10^{-4}$ & $1 + 0.00456 \left(c_{Hl}^{(1)}\right)_{33} + 0.200 \left(c_{Hl}^{(3)}\right)_{33} - 0.0320 \left(c_{He}\right)_{33}$ \\
\hline
\multirow{2}{*}{$[800, 1000)$} & $\leq 0.5$ & $2.83 \times 10^{-5}$ & $1 - 0.0291 \left(c_{Hl}^{(1)}\right)_{33} + 0.358 \left(c_{Hl}^{(3)}\right)_{33} - 0.0726 \left(c_{He}\right)_{33}$ \\
\cline{2-4}
& $> 0.5$ & $7.42 \times 10^{-5}$ & $1 - 0.0141 \left(c_{Hl}^{(1)}\right)_{33} + 0.213 \left(c_{Hl}^{(3)}\right)_{33} - 0.0362 \left(c_{He}\right)_{33}$ \\
\hline
\multirow{2}{*}{$[1000, 1500)$} & $\leq 0.6$ & $4.84 \times 10^{-5}$ & $1 - 0.0712 \left(c_{Hl}^{(1)}\right)_{33} + 0.376 \left(c_{Hl}^{(3)}\right)_{33} - 0.0823 \left(c_{He}\right)_{33}$ \\
\cline{2-4}
& $> 0.6$ & $7.97 \times 10^{-5}$ & $1 - 0.0329 \left(c_{Hl}^{(1)}\right)_{33} + 0.227 \left(c_{Hl}^{(3)}\right)_{33} - 0.0408 \left(c_{He}\right)_{33}$ \\
\hline
\multirow{2}{*}{$[1500, 2000)$} & $\leq 0.7$ & $2.49 \times 10^{-5}$ & $1 - 0.131 \left(c_{Hl}^{(1)}\right)_{33} + 0.441 \left(c_{Hl}^{(3)}\right)_{33} - 0.103 \left(c_{He}\right)_{33}$ \\
\cline{2-4}
& $> 0.7$ & $3.04 \times 10^{-5}$ & $1 - 0.0524 \left(c_{Hl}^{(1)}\right)_{33} + 0.231 \left(c_{Hl}^{(3)}\right)_{33} - 0.0453 \left(c_{He}\right)_{33}$ \\
\hline
\multirow{2}{*}{$[2000, +\infty)$} & $\leq 0.75$ & $3.30 \times 10^{-5}$ & $1 - 0.200 \left(c_{Hl}^{(1)}\right)_{33} + 0.539 \left(c_{Hl}^{(3)}\right)_{33} - 0.128 \left(c_{He}\right)_{33}$ \\
\cline{2-4}
& $> 0.75$ & $4.38 \times 10^{-5}$ & $1 - 0.0679 \left(c_{Hl}^{(1)}\right)_{33} + 0.224 \left(c_{Hl}^{(3)}\right)_{33} - 0.0492 \left(c_{He}\right)_{33}$ \\
\midrule
\midrule
$\mu^{-}\mu^{+}\rightarrow \tau \nu\bar{\tau} \nu_{\mu} \mu$ & $-$ & 0.351 & $1 - 6.83 \times 10^{-5} \left(c_{Hl}^{(1)}\right)_{33} + 0.122 \left(c_{Hl}^{(3)}\right)_{33} - 5.55 \times 10^{-8} \left(c_{He}\right)_{33}$ \\
\bottomrule
\end{tabular}
}
\caption{The SM cross section and normalized SMEFT cross section ($\sigma_{\rm SMEFT}/\sigma_{\rm SM}$, as a function of Wilson coefficients) of $\mu^-\mu^+ \to \tau^- \tau^+\nu_{\mu}\bar{\nu}_{\mu}$ for different invariant mass and $\cos\bar{\theta}$ bins at $\sqrt{s}=10\TeV$. The first (last) row corresponds to the unbinned total cross section for $\mu^-\mu^+ \to \tau^- \tau^+\nu_{\mu}\bar{\nu}_{\mu}$ ($\mu^-\mu^+ \to \tau \nu\bar{\tau} \nu_{\mu} \mu$). Tagging efficiencies are not applied here.}
\label{tab:xs_tata10}
\end{table}

\begin{table}[h]
\centering
\renewcommand{\arraystretch}{1.5}
\resizebox{\textwidth}{!}{
\begin{tabular}{m{4cm}<{\centering}|m{2cm}<{\centering}|m{2cm}<{\centering}|m{10cm}<{\centering}}
\toprule
Invariant mass [GeV]\vspace{4pt} & Polar angle $\cosbt$ & SM cross section [pb] & Normalized\vspace{4pt} SMEFT cross section \\
\midrule
$[0, +\infty)$ & -- & 0.0800 & $1 + 0.126 \left(c_{Hl}^{(1)}\right)_{33} + 0.131 \left(c_{Hl}^{(3)}\right)_{33} - 0.102 \left(c_{He}\right)_{33}$ \\
\midrule
\midrule
$[0, 100)$ & -- & 0.0721 & $1 + 0.129 \left(c_{Hl}^{(1)}\right)_{33} + 0.130 \left(c_{Hl}^{(3)}\right)_{33} - 0.102 \left(c_{He}\right)_{33}$ \\
\hline
$[100, 200)$ & -- & 0.00464 & $1 + 0.141 \left(c_{Hl}^{(1)}\right)_{33} + 0.145 \left(c_{Hl}^{(3)}\right)_{33} - 0.0626 \left(c_{He}\right)_{33}$ \\
\hline
$[200, 400)$ & -- & $0.00102$ & $1 + 0.0792 \left(c_{Hl}^{(1)}\right)_{33} + 0.166 \left(c_{Hl}^{(3)}\right)_{33} - 0.0267 \left(c_{He}\right)_{33}$ \\
\hline
$[400, 600)$ & -- & $3.73 \times 10^{-4}$ & $1 + 0.0346 \left(c_{Hl}^{(1)}\right)_{33} + 0.207 \left(c_{Hl}^{(3)}\right)_{33} - 0.0349 \left(c_{He}\right)_{33}$ \\
\hline
\multirow{2}{*}{$[600, 800)$} & $\leq 0.5$ & $6.73 \times 10^{-5}$ & $1 + 0.000268 \left(c_{Hl}^{(1)}\right)_{33} + 0.300 \left(c_{Hl}^{(3)}\right)_{33} - 0.0617 \left(c_{He}\right)_{33}$ \\
\cline{2-4}
 & $> 0.5$ & $1.32 \times 10^{-4}$ & $1 + 0.000162 \left(c_{Hl}^{(1)}\right)_{33} + 0.203 \left(c_{Hl}^{(3)}\right)_{33} - 0.0365 \left(c_{He}\right)_{33}$ \\
\hline
\multirow{2}{*}{$[800, 1000)$} & $\leq 0.5$ & $3.71 \times 10^{-5}$ & $1 - 0.0469 \left(c_{Hl}^{(1)}\right)_{33} + 0.372 \left(c_{Hl}^{(3)}\right)_{33} - 0.0864 \left(c_{He}\right)_{33}$ \\
\cline{2-4}
 & $> 0.5$ & $8.31 \times 10^{-5}$ & $1 - 0.0245 \left(c_{Hl}^{(1)}\right)_{33} + 0.222 \left(c_{Hl}^{(3)}\right)_{33} - 0.0435 \left(c_{He}\right)_{33}$ \\
\hline
\multirow{2}{*}{$[1000, 1500)$} & $\leq 0.55$ & $5.56 \times 10^{-5}$ & $1 - 0.124 \left(c_{Hl}^{(1)}\right)_{33} + 0.448 \left(c_{Hl}^{(3)}\right)_{33} - 0.116 \left(c_{He}\right)_{33}$ \\
\cline{2-4}
 & $> 0.55$ & $1.01 \times 10^{-4}$ & $1 - 0.0564 \left(c_{Hl}^{(1)}\right)_{33} + 0.251 \left(c_{Hl}^{(3)}\right)_{33} - 0.0539 \left(c_{He}\right)_{33}$ \\
\hline
\multirow{2}{*}{$[1500, 2000)$} & $\leq 0.5$ & $1.85 \times 10^{-5}$ & $1 - 0.301 \left(c_{Hl}^{(1)}\right)_{33} + 0.702 \left(c_{Hl}^{(3)}\right)_{33} - 0.222 \left(c_{He}\right)_{33}$ \\
\cline{2-4}
 & $> 0.5$ & $5.32 \times 10^{-5}$ & $1 - 0.116 \left(c_{Hl}^{(1)}\right)_{33} + 0.300 \left(c_{Hl}^{(3)}\right)_{33} - 0.0723 \left(c_{He}\right)_{33}$ \\
\hline
\multirow{2}{*}{$[2000, +\infty)$} & $\leq 0.6$ & $4.35 \times 10^{-5}$ & $1 - 0.727 \left(c_{Hl}^{(1)}\right)_{33} + 1.09 \left(c_{Hl}^{(3)}\right)_{33} - 0.444 \left(c_{He}\right)_{33}$ \\
\cline{2-4}
 & $> 0.6$ & $8.61 \times 10^{-5}$ & $1 - 0.283 \left(c_{Hl}^{(1)}\right)_{33} + 0.457 \left(c_{Hl}^{(3)}\right)_{33} - 0.113 \left(c_{He}\right)_{33}$ \\
\midrule
\midrule
$\mu^{-}\mu^{+}\rightarrow \tau \nu_\tau \nu_{\mu} \mu$ & $-$ & 0.248 & $1 - 9.16 \times 10^{-5} \left(c_{Hl}^{(1)}\right)_{33} + 0.121 \left(c_{Hl}^{(3)}\right)_{33} - 8.98 \times 10^{-8} \left(c_{He}\right)_{33}$ \\
\bottomrule
\end{tabular}
}
\caption{Same as \autoref{tab:xs_tata10} but for $\sqrt{s}=30\TeV$.}
\label{tab:xs_tata30}
\end{table}

\clearpage

\section{Additional results}
\label{app:delta_rho}

In this appendix, we illustrate the contours for other different fermions production processes mentioned in \autoref{subsec:invbin}, \ref{subsec:cosbin} and \ref{subsec:unpair}.

\begin{figure}[htb]
\centering
\subfigure{
    \includegraphics[width=0.30\textwidth]{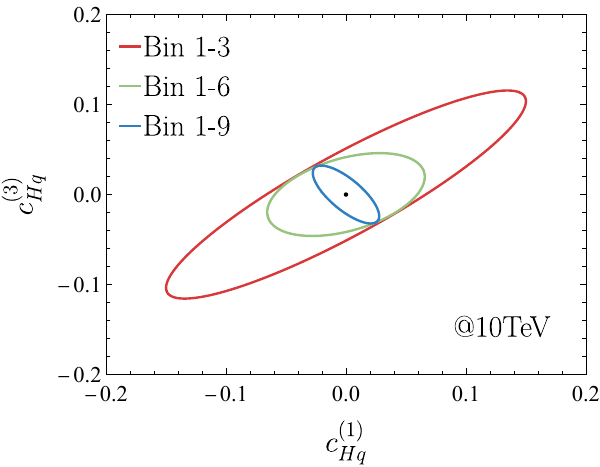}
    \label{subfig:chq1chq3_10TeV_cc}
}
\subfigure{
    \includegraphics[width=0.30\textwidth]{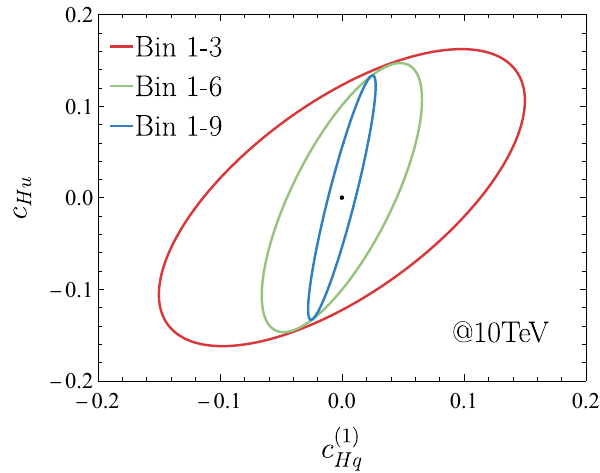}
    \label{subfig:chq1chu_10TeV_cc}
}
\subfigure{
    \includegraphics[width=0.30\textwidth]{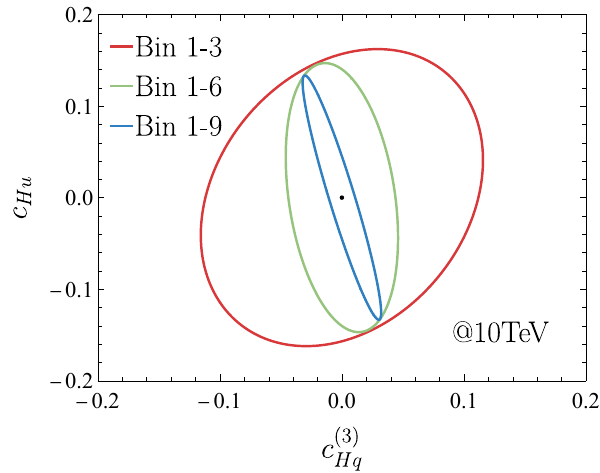}
    \label{subfig:chq3chu_10TeV_cc}
}
\newline
\subfigure{
    \includegraphics[width=0.30\textwidth]{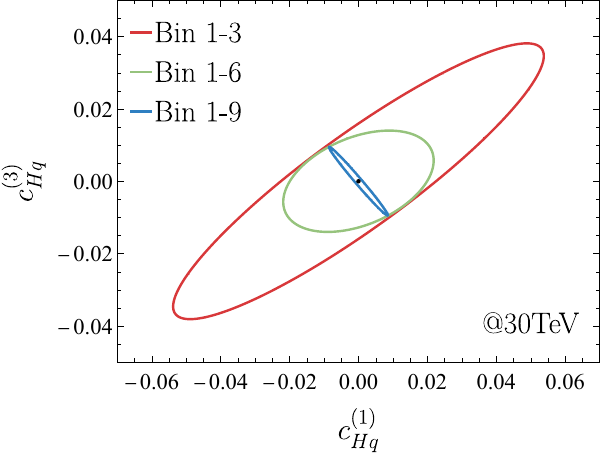}
    \label{subfig:chq1chq3_30TeV_cc}
}
\subfigure{
    \includegraphics[width=0.30\textwidth]{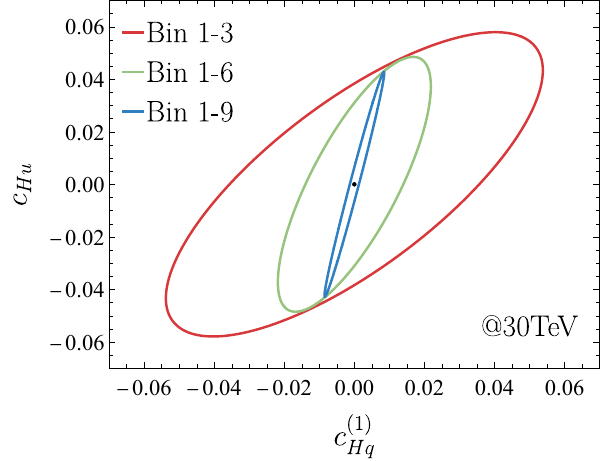}
    \label{subfig:chq1chu_30TeV_cc}
}
\subfigure{
    \includegraphics[width=0.30\textwidth]{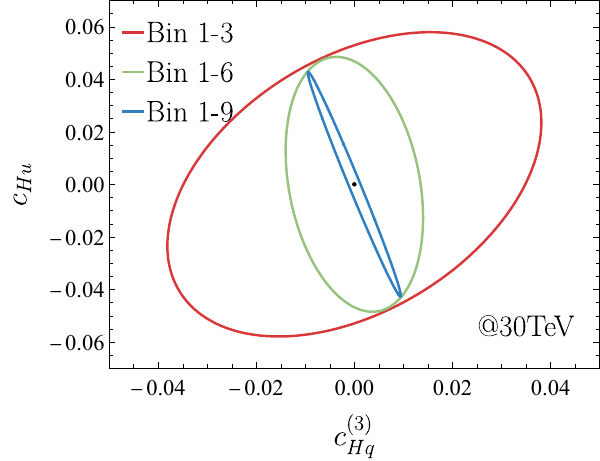}
    \label{subfig:chq3chu_30TeV_cc}
}
\caption{Same as \autoref{fig:inv_bb}, the $\Delta\chi^2=1$ contours from the 3-parameter $(c_{Hq}^{(1)},c_{Hq}^{(3)},c_{Hu})$ but fit to $\mu^{-}\mu^{+}\rightarrow c\bar{c}\nu_{\mu}\bar{\nu}_{\mu}$ at 10\,TeV (top row) and 30\,TeV (bottom row), with the invariant mass bins in \autoref{tab:invbin}.  For each row, the results are projected onto three 2D planes (each with the other parameter marginalized). Only the contours with the first 3, 6, and all 9 bins are shown.
}
\label{fig:inv_cc}
\end{figure}

\begin{figure}[htb]
\centering
\hspace{0.1cm}
\subfigure{
    \includegraphics[width=0.30\textwidth]{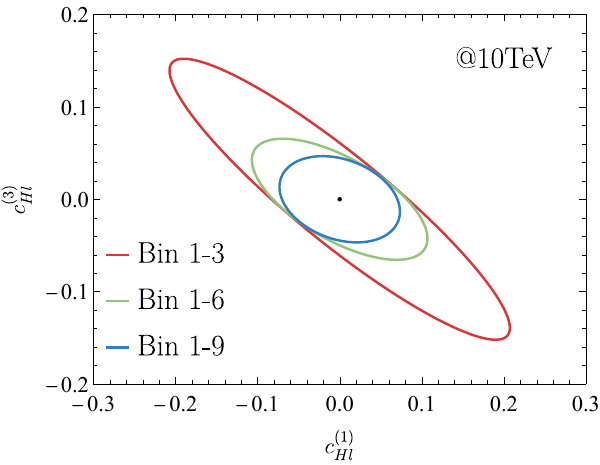}
    \label{subfig:chl1chl3_10TeV_tata}
}
\hspace{-0.3cm}
\subfigure{
    \includegraphics[width=0.30\textwidth]{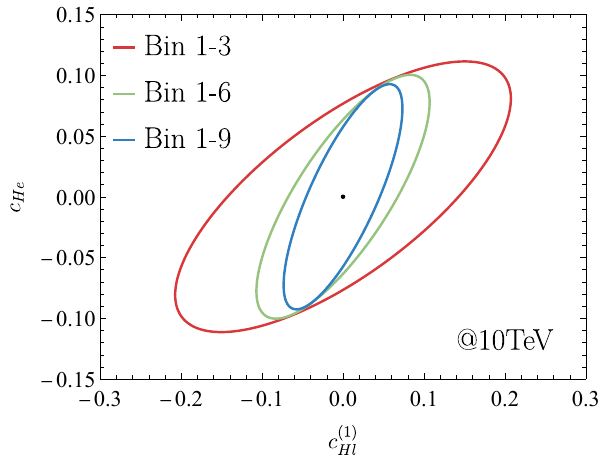}
    \label{subfig:chl1che_10TeV_tata}
}
\subfigure{
    \includegraphics[width=0.30\textwidth]{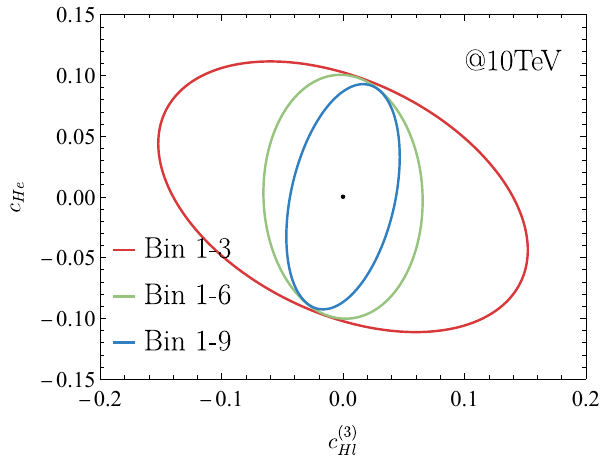}
    \label{subfig:chl3che_10TeV_tata}
}
\newline
\subfigure{
    \includegraphics[width=0.30\textwidth]{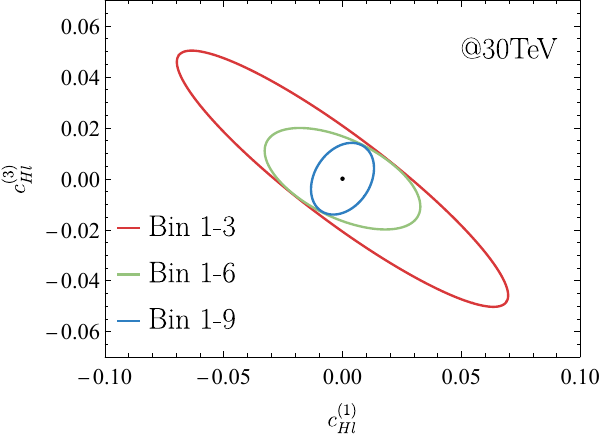}
    \label{subfig:chl1chl3_30TeV_tata}
}
\subfigure{
    \includegraphics[width=0.30\textwidth]{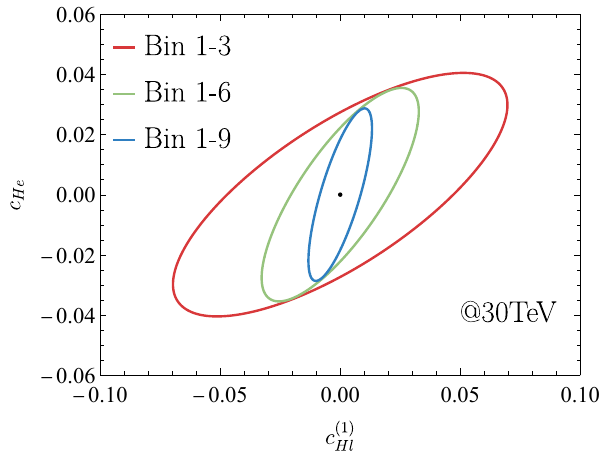}
    \label{subfig:chl1che_30TeV_tata}
}
\subfigure{
    \includegraphics[width=0.30\textwidth]{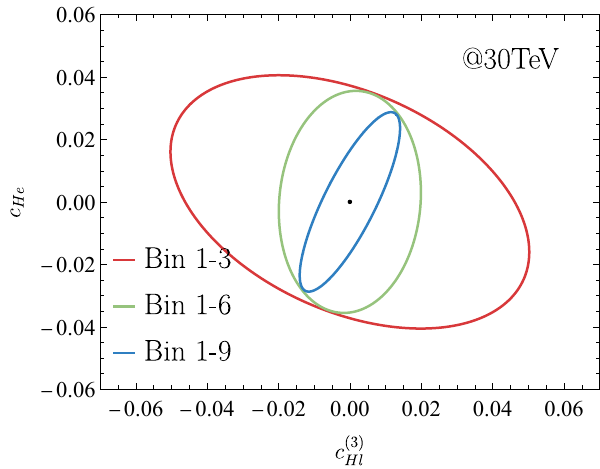}
    \label{subfig:chl3che_30TeV_tata}
}
\caption{Same as \autoref{fig:inv_cc} but for $\mu^{-}\mu^{+}\rightarrow \tau^{-}\tau^{+}\nu_{\mu}\bar{\nu}_{\mu}$ with Wilson coefficients $c^{(1)}_{Hl}$, $c^{(3)}_{Hl}$ and $c_{He}$. 
}
\label{fig:inv_tata}
\end{figure}

\begin{figure}[htb]
\centering
\subfigure{
    \includegraphics[width=0.3\textwidth]{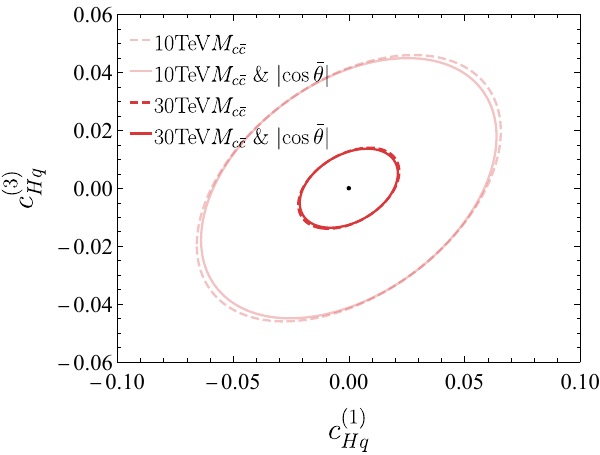}
    \label{subfig:chq1chq3_cc}
}
\subfigure{
    \includegraphics[width=0.3\textwidth]{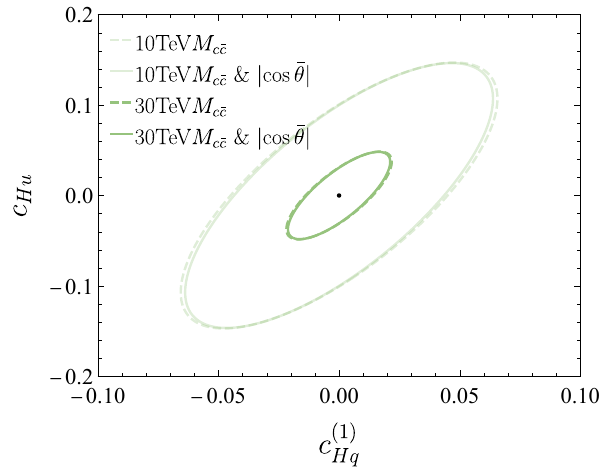}
    \label{subfig:chq1chu_cc}
}
\subfigure{
    \includegraphics[width=0.3\textwidth]{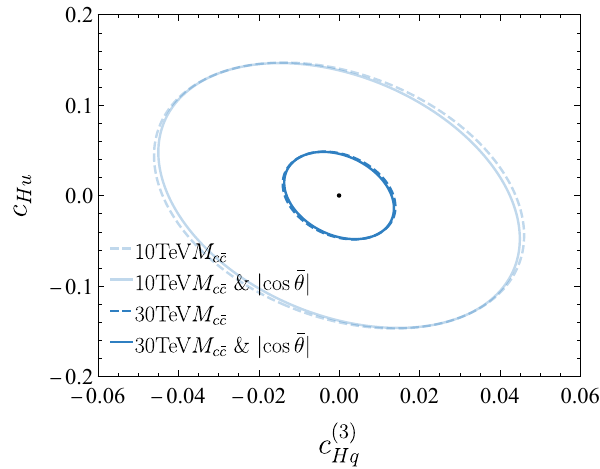}
    \label{subfig:chq3chu_cc}
}
\caption{Same as \autoref{fig:cosbin_bb}, comparison of the results but for  $\mu^{-}\mu^{+} \to c\bar{c}\nu_{\mu}\bar{\nu}_{\mu}$ with only the invariant mass bins (dashed contours, binning as in \autoref{tab:invbin}) and also with the $\cosbt$ bins (solid contours, binning as in \autoref{tab:divide}). An invariant mass cut of $M_{c\bar{c}}<1$\,TeV is imposed.  The contours correspond to $\Delta\chi^2=1$ from the 3-parameter $(c_{Hq}^{(1)},c_{Hq}^{(3)},c_{Hu})$ fit. 
}
\label{fig:cosbin_cc}
\end{figure}

\begin{figure}[htb]
\centering
\subfigure{
    \includegraphics[width=0.30\textwidth]{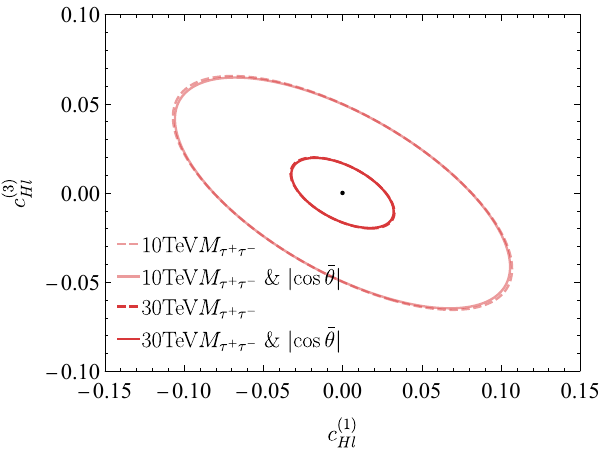}
    \label{subfig:chl1chl3_tata}
}
\subfigure{
    \includegraphics[width=0.30\textwidth]{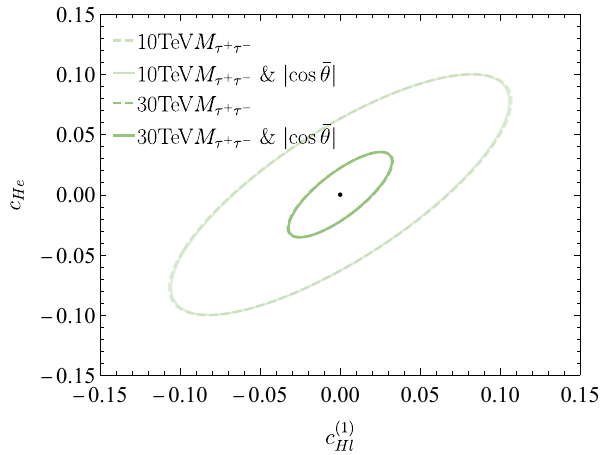}
    \label{subfig:chl1che_tata}
}
\subfigure{
    \includegraphics[width=0.30\textwidth]{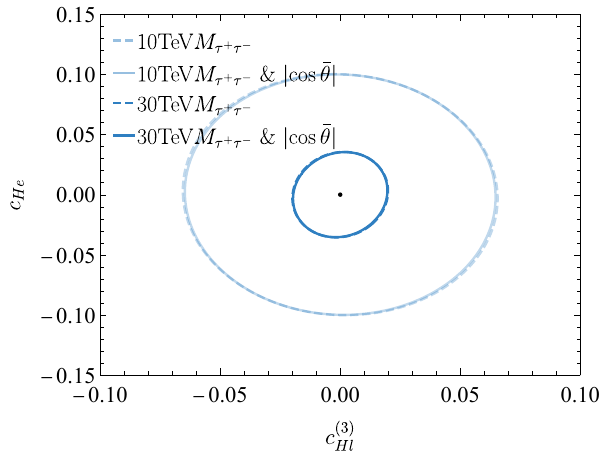}
    \label{subfig:chl3che_tata}
}
\caption{Same as \autoref{fig:cosbin_cc} but for $\mu^{-}\mu^{+}\rightarrow \tau^-\tau^+ \nu_{\mu}\bar{\nu}_{\mu}$ with Wilson coefficients $c^{(1)}_{Hl}$, $c^{(3)}_{Hl}$ and $c_{He}$.
}
\label{fig:cosbin_tata}
\end{figure}

\begin{figure}[htb]
\centering
\subfigure{
    \includegraphics[width=0.30\textwidth]{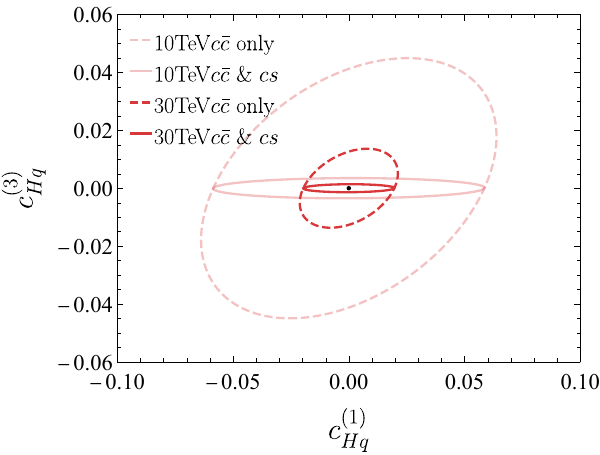}
    \label{subfig:chq1chq3_cc_plus}
}
\subfigure{
    \includegraphics[width=0.30\textwidth]{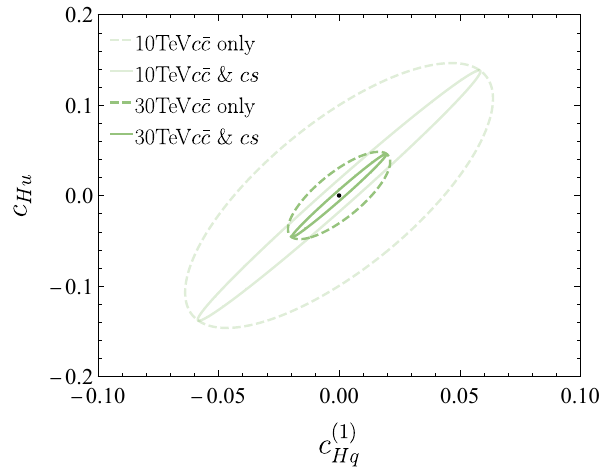}
    \label{subfig:chq1chu_cc_plus}
}
\subfigure{
    \includegraphics[width=0.30\textwidth]{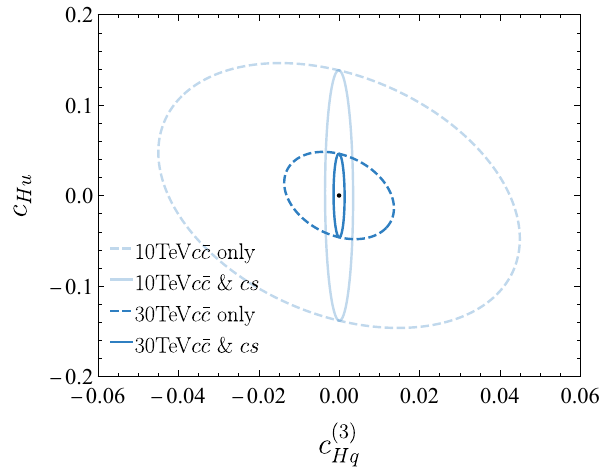}
    \label{subfig:chq3chu_cc_plus}
}
\caption{Same as \autoref{fig:unpair_tata}, comparison of the results for the three parameters $c_{Hq}^{(1)}, c_{Hq}^{(3)}, c_{Hu}$ from the measurements of $\mu^-\mu^+ \to c\bar{c} \nu_{\mu}\bar{\nu}_{\mu}$ only (labeled as ``$c\bar{c}$ only'', dashed contours, with $M_{c\bar{c}}<1\,$TeV) and the combination of the $\mu^-\mu^+ \to c \bar{c}\nu_{\mu}\bar{\nu}_{\mu}$ and $\mu^-\mu^+ \to c s \nu_{\mu} \mu$ (labeled as ``$c\bar{c} \& c s$'', solid contours).  Contours corresponds to $\Delta\chi^2=1$. 
}
\label{fig:unpair_cc}
\end{figure}

\autoref{fig:inv_cc} and \autoref{fig:inv_tata} show the $\Delta\chi^2=1$ contours with invariant mass bin, same as \autoref{fig:inv_bb}, but for $\mu^{-}\mu^{+}\rightarrow c\bar{c}\nu_{\mu}\bar{\nu}_{\mu}$ and $\mu^{-}\mu^{+}\rightarrow \tau^{-}\tau^{+}\nu_{\mu}\bar{\nu}_{\mu}$, respectively. 
\autoref{fig:cosbin_cc} and \autoref{fig:cosbin_tata} show the comparison of the results with and without including the $\cosbt$ bins, same as \autoref{fig:cosbin_bb}, but for $\mu^{-}\mu^{+}\rightarrow c\bar{c}\nu_{\mu}\bar{\nu}_{\mu}$ and $\mu^{-}\mu^{+}\rightarrow \tau^{-}\tau^{+}\nu_{\mu}\bar{\nu}_{\mu}$, respectively. 
\autoref{fig:unpair_cc} shows the comparison of asymmetric processes with and without included, same as \autoref{fig:unpair_tata}, but for the processes $\mu^-\mu^+ \to c \bar{c}\nu_{\mu}\bar{\nu}_{\mu}$ and $\mu^-\mu^+ \to c s \nu_{\mu} \mu$.

\newpage
\bibliographystyle{JHEP}
\bibliography{paper.bib}


\end{document}